\documentclass[12pt,preprint]{aastex}

\newbox\grsign \setbox\grsign=\hbox{$>$} \newdimen\grdimen \grdimen=\ht\grsign
\newbox\simlessbox \newbox\simgreatbox
\setbox\simgreatbox=\hbox{\raise.5ex\hbox{$>$}\llap
     {\lower.5ex\hbox{$\sim$}}}\ht1=\grdimen\dp1=0pt
\setbox\simlessbox=\hbox{\raise.5ex\hbox{$<$}\llap
     {\lower.5ex\hbox{$\sim$}}}\ht2=\grdimen\dp2=0pt
\def\simgreat{\mathrel{\copy\simgreatbox}}
\def\simless{\mathrel{\copy\simlessbox}}
\newbox\simppropto
\setbox\simppropto=\hbox{\raise.5ex\hbox{$\sim$}\llap
     {\lower.5ex\hbox{$\propto$}}}\ht2=\grdimen\dp2=0pt

\shorttitle{EZ\_Ages Algorithm}
\shortauthors{Graves \& Schiavon}

\begin{document}

\title{Measuring Ages and Elemental Abundances from Unresolved Stellar
  Populations: Fe, Mg, C, N, and Ca}

\author{Genevieve J. Graves\altaffilmark{1} \& Ricardo P. Schiavon\altaffilmark{2}}

\altaffiltext{1}{UCO/Lick Observatory, University of California, Santa
	Cruz, CA 95064}
\altaffiltext{2}{Gemini Observatory, 670 N. A'ohoku Place, Hilo, HI 96720}

\begin{abstract}

We present a method for determining mean light-weighted ages and
abundances of Fe, Mg, C, N, and Ca, from medium resolution
spectroscopy of unresolved stellar populations. The method, pioneered
by Schiavon (2007), is implemented in a publicly available code called
EZ\_Ages. The method and error estimation are described, and the
results tested for accuracy and consistency, by application to
integrated spectra of well-known Galactic globular and open
clusters. Ages and abundances from integrated light analysis agree
with studies of resolved stars to within $\pm0.1$dex for most
clusters, and to within $\pm0.2$ dex for nearly all cases. The results
are robust to the choice of Lick indices used in the fitting to within
$\pm0.1$ dex, except for a few systematic deviations which are clearly
categorized. The realism of our error estimates is checked through
comparison with detailed Monte Carlo simulations. Finally, we apply
EZ\_Ages to the sample of galaxies presented in \citet{tho05} and
compare our derived values of age, [Fe/H], and [$\alpha$/Fe] to their
analysis. We find that [$\alpha$/Fe] is very consistent between the
two analyses, that ages are consistent for old ($Age > 10$ Gyr)
populations, but show modest systematic differences at younger ages,
and that [Fe/H] is fairly consistent, with small systematic
differences related to the age systematics. Overall, EZ\_Ages provides
accurate estimates of fundamental parameters from medium resolution
spectra of unresolved stellar populations in the old and
intermediate-age regime, for the first time allowing quantitative
estimates of the abundances of C, N, and Ca in these unresolved
systems.

\end{abstract}


\section{Introduction}

The study of stellar populations in galaxies is going through very
exciting transformations.  The advent of extensive and high-quality
data sets for galaxies in the local and distant universe poses a
demand for ever improving stellar population synthesis models.  In
this paper, we present a new tool, named EZ\_Ages, that simultaneously
address two on-going needs in the field of stellar population
synthesis: 1) it extracts accurate ages and abundances of several
elements from integrated spectra of stellar populations, and 2) it
performs this task in an automatic fashion, which is well-suited for
applications to large data sets.

Early stellar population synthesis models for single-burst
populations---so called ``single stellar populations'' or SSPs---were
based on low resolution stellar spectra.  The models of \citet{wor94}
and \citet{vaz96} used optical absorption-line indices with 8--10
{\AA} resolution and limited spectral coverage, while \citet{bc93}
synthesized model spectra with broad wavelength coverage (ultraviolet
through infrared) and resolution of 10--20 {\AA} in the optical.  All
models were for solar-abundance ratios only.  Since then, a new
generation of SSP models have pushed to higher resolution (2--3 {\AA}
for \citealt{bc03} and \citealt{vaz03}) and to variable abundance
patterns (\citealt{bor95}; \citealt{wei95}; \citealt{tan98};
\citealt{tra00}; \citealt{tho03}; \citealt{mar05}; \citealt{sch07} and
\citealt{coe07}).

It is well known that many stellar populations, including the Galactic
halo and bulge, Galactic globular clusters, massive early-type
galaxies, and the bulges of spiral galaxies, have super-solar
[$\alpha$/Fe]. In resolved systems, where abundance analysis of
individual stars is possible, it has been known for a while that not
all $\alpha$ elements are equally enhanced.  On the other hand, in
unresolved systems it has been only recently that evidence started
accumulating to the fact that observed stellar populations have
abundance patterns which cannot be described by only two parameters
(ie.  [Fe/H] and [$\alpha$/Fe]) but rather require more complicated
abundance patterns.  In particular, it seems that not all the $\alpha$
elements are equally enhanced in early-type galaxies (e.g.,
\citealt{vaz97}; \citealt{wor98}; \citealt{tra98}; \citealt{hen99};
\citealt{tho03}; \citealt{sch07}).  In the best studied, closest,
resolved spheroidal system, the Galactic bulge, there is also recent
evidence that not all $\alpha$-elements are equally enhanced
\citep{ful05,ful07}.  In particular, \citet{ful07} found that, while
Mg is strongly enhanced relative to Fe ([Mg/Fe] $\sim$ +0.3 across a
range of [Fe/H]), other $\alpha$ elements such as Si, Ca, Ti, and O
tend to follow a disk-like trend, with [X/Fe] decreasing with
increasing [Fe/H].  Interestingly, in the case of oxygen---arguably
the most important among the $\alpha$ elements---they found [O/Fe]
$\approx 0$ at [Fe/H] = 0.  \citet{ben04} show a similar result for
Milky Way disk stars in both the thin and thick disk.  \citet{ath03}
demonstrates that warm ionized gas in a sample of 7 local early-type
galaxies has mean [O/H] $\approx -0.1$, while the stellar populations
of the galaxies have super-solar [Mg/H].  Clues to variations of the
abundances of other elements, such as C and N, particularly as a
function of environment, have been found in a number of studies.  The
Lick indices C$_2$4668 and CN$_2$, which are sensitive to C and O (and
N, in the case of CN$_2$), have been shown to be weaker in early-type
galaxies in dense environments as compared to galaxies in the field,
while the strength of the Mg~{\it b} index remains unchanged across
environments \citep{san03}.  Finally, a number of studies have found
evidence that Ca seems to behave as an Fe-group element in elliptical
galaxies, rather than following the trend of other $\alpha$ elements
(\citealt{vaz97}; \citealt{wor98}; \citealt{tra98}; \citealt{hen99};
\citealt{tho03}).  This result has recently been questioned by
\citet{pro05} and \citet{sch07}, who show that the Ca4227 index is
strongly affected by CN lines, which may be partially responsible for
the low [Ca/Fe] suggested by other authors.  Although
$\alpha$-enhanced models are an important improvement over
solar-scaled-only, it is fundamental that models with more degrees of
freedom be developed, which are able to explore the complexity
suggested by the data.  By establishing a more detailed snapshot of
the abundance patterns of stars in galaxies, these models can pose
fundamental constraints on models of galaxy formation.

\citet{sch07} presents a new set of SSP models which do exactly that.
These models explicitly include adjustable abundance patterns for
multiple elements, allowing the user to separately vary the abundances
of C, N, O, Mg, Ca, Na, Si, Cr, and Ti.  This paper describes an
algorithm for searching through the space of variable abundance models
to find the abundance mix which best fits the Lick indices measured in
a given input stellar population spectrum.  This algorithm has been
implemented in an IDL code package, called ``EZ\_Ages'' which is
publicly available for use.  Section \ref{seq_grid} describes the
sequential grid inversion algorithm.  In \S\ref{compare_clusters}, we
present ages and abundances estimated by EZ\_Ages for the Galactic
globular clusters NGC 6121, 47 Tuc, NGC 6441, and NGC 6528, as well as
the open cluster M67.  These values are compared to results in the
literature determined from isochrone fitting of cluster photometry and
from high-resolution spectral abundance analysis for individual stars
in the clusters.  In \S\ref{compare_galaxies} we apply EZ\_Ages to
galaxy data from the literature, contrasting our results with those by
other authors.  Section \ref{conc} summarizes our conclusions.

The beta-version of EZ\_Ages is available for download, along with
instructions on installation and
use.\footnote[1]{www.ucolick.org/$\sim$graves/EZ\_Ages.html}

\section{Sequential Grid Inversion Algorithm}\label{seq_grid}

\subsection{Schiavon Simple Stellar Population Models}

The code EZ\_Ages is designed for use with the single stellar
population models described in \citet[hereafter S07]{sch07}.  The
details of the model are described in S07; here we briefly discuss the
aspects of the model relevant to its practical use.

These models include a choice of two sets of isochrones by the Padova
group: one is the solar-scaled isochrones of \citet{gir00} and the
other is the $\alpha$-enhanced set of isochrones by \citet{sal00} with
average [$\alpha/$Fe$] = +0.42$ and with individual abundance ratios
chosen to match metal-poor field stars.\footnote[2]{The following
$\alpha$ elements are enhanced as shown: [O/Fe] = $+0.50$, [Ne/Fe] =
$+0.29$, [Mg/Fe] = $+0.40$, [Si/Fe] = $+0.30$, [S/Fe] = $+0.33$,
[Ca/Fe] = $+0.50$, [Ti/Fe] = $+0.63$, [Ni/Fe] = $+0.02$.  All other
abundance-ratios are solar.}  However, the models based on the
$\alpha$-enhanced isochrone must be used with caution because
\citet{wei06} have shown that, due to a problem with the opacity
tables adopted, those isochrones predict temperatures that are
slightly too high at both the main-sequence turnoff and the giant
branch.  The effect is strongest at solar metallicity, where it
results in ages that are slightly too old.  For details, see \S4.3 of
S07.  Developing isochrones with variable abundance ratios is an
active area of current research (see recent work by \citealt{dot07})
and we hope to incorporate a broader range of isochrones into EZ\_Ages
in the near future.  

In addition to a choice of isochrones, the S07 models allow the user
to select an input abundance pattern.  The abundance ratios that can
be varied are: [Mg/Fe], [C/Fe], [N/Fe], [O/Fe], [Ca/Fe], [Na/Fe],
[Si/Fe], [Cr/Fe], and [Ti/Fe].  The S07 models output predicted Lick
index measurements for a grid of age and [Fe/H] values, based upon the
input abundance ratios.  The models span the age range $0.1 \le t \le
15.8$ Gyr, with $-1.3 \le$ [Fe/H] $\le +0.2$ for the solar-scaled
isochrone and $-0.8 \le$ [Fe/H] $\le +0.3$ for the $\alpha$-enhanced
isochrone, though only models older than 0.8 Gyr are computed with
variable abundance ratios.

The Lick indices included in the S07 model are H$\delta_A$,
H$\delta_F$, H$\gamma_A$, H$\gamma_F$, H$\beta$, CN$_1$, CN$_2$,
Ca4227, G4300, Fe4383, C$_2$4668, Fe5015, Mg$_2$, Mg~{\it b}, Fe5270,
and Fe5335, as defined in Table 1 of \citet{wor94} and Table 1 of
\citet{wor97}.  The output model indices can then be compared with
Lick indices measured in the integrated spectrum of a real stellar
population (i.e., a stellar cluster or a galaxy) to determine the
best-fitting mean light-weighted age and [Fe/H] for the stellar
population, for a given set of model abundance ratios.

The Lick indices included in the S07 model are summarized in Table
\ref{indextable}, along with their main abundance sensitivities.
These are taken from the works of \citet{ser05} and \citet{kor05}.
The sensitivities from \citet{kor05} are taken from their Tables 4 and
5, which give abundance sensitivities for turnoff stars and giant
branch stars respectively.  We have chosen to include only those two
evolutionary phases in the table, as they dominate the stellar
population model spectrum (accounting for $\sim 90$\% of the
integrated light), even though all three evolutionary phases are taken
into account in the model computations.  In addition to the
sensitivities to individual element abundances shown in the table, all
indices are sensitive to changes in the total metallicity.  We
interpret the effect of [Z/H] on Fe5270 and Fe5335 as being due
primarily to changes in Fe because none of the other elements
investigated by KMT05 appear to affect these indices.  Table
\ref{indextable} shows reasonable agreement between the two different
sets of sensitivity tables.

Abundance variations enter the model in two ways, firstly through the
choice of isochrone (solar-scaled or $\alpha$-enhanced).  At this
stage, the model takes into account the effect of the abundance mix on
stellar evolution---specifically, how the chemical mix affects
temperatures and luminosities of stars of different masses and at
different evolutionary stages.  Admittedly, this is done in an
approximate fashion, given that unfortunately we only have isochrones
computed for two abundance patterns.  Secondly, the effect of
individual elemental abundances on line and continuum opacities in
stellar atmospheres is incorporated into model Lick index predictions,
using the response functions computed by \citet{kor05}.  At this
second stage, it is possible to vary elemental abundances
independently; at the first stage, the only choices are
[$\alpha$/Fe]~=~0 and $+0.42$, with total metallicity varying.

It should be stated clearly that these models are cast in terms of
[Fe/H], and not total metallicity, [Z/H].  This reflects our choice to
deal explicitly with quantities that can be inferred from measurements
taken in the integrated spectra of galaxies---total metallicity not
being one of them, given our current inability to use integrated
spectra of stellar populations to constrain the most abundant of all
metals, oxygen (see discussion in \S4.4.1 of S07).  Another advantage
with casting models in terms of [Fe/H] is that each elemental
abundance can be treated separately, so that the effect of its
variation can be studied in isolation from every other elemental
abundance (at the cost of varying the total metallicity).  In the case
of models cast in terms of [Z/H], it is impossible to vary the
abundance of a single element, because enhancing one element means
decreasing the abundances of all other elements to keep total
metallicity constant.  As an example, in models cast in terms of
[Z/H], a solar-metallicity ([Z/H]~=~0) $\alpha$-enhanced population
has only a slightly higher abundance of $\alpha$ elements than the
Sun, while having a much lower iron-abundance.  It is the large iron
abundance difference that is responsible for the bulk of the
difference between those two hypothetical models, {\it not} the small
difference in $\alpha$-element abundances.  For this reason, claims
that higher order Balmer lines are strongly affected by
$\alpha$-element abundances \citep{tho04} should be taken with caution
(see \S4.3 S07 for a thorough discussion of this point).

\subsection{Fiducial Age and [Fe/H]}

In practice, one would like to use absorption line measurements
in the integrated spectrum of a given stellar population to determine
not only the age and [Fe/H] for a fixed set of abundance ratios,
but to find the best set of input abundances as well.  The brute-force
method of finding the best-fitting abundances would involve generating
a set of models that span the available space of age, [Fe/H], and
abundance patterns, and then identifying the model whose predicted
Lick indices best match those measured in the spectrum.  However,
with nine variable abundance ratios, the search space quickly becomes
very large.  If each of those abundance ratios is represented by
only four possible values, with 23 model ages and four possible
values of [Fe/H], that would mean creating more than 24 million
models!

Instead, we choose to perform a directed search for the best model,
taking advantage of the fact that various Lick indices are sensitive
to only a few, often different, elemental abundances, as indicated in
Table~\ref{indextable}.  These sensitivity variations have been
modeled by several groups using spectrum synthesis, based on model
stellar atmospheres (see recent papers by \citealt{kor05} and
\citealt{ser05}).  A more detailed discussion of the motivation behind
this method can be found in S07.

EZ\_Ages begins by computing a set of models with the chosen isochrone
(either solar-scaled or $\alpha$-enhanced), using solar abundance
ratios for the stellar atmospheres.  It then uses a pair of lines that
are sensitive to age and [Fe/H], but relatively {\it insensitive} to
other elemental abundances---a Balmer line and an Fe line---to
determine a fiducial age and [Fe/H] for the stellar population in
question.  From Table 1 we can see that H$\beta$ and the iron lines
Fe5270 and Fe5335 are good choices for determining the fiducial, as
they are mostly insensitive to other element abundances.  In EZ\_Ages,
the default choice of indices for calculating fiducial age and [Fe/H]
are H$\beta$ and $\langle$Fe$\rangle$, an average of Fe5270 and
Fe5335.

The top left panel of Figure \ref{grids} shows a plot of model grids
for H$\beta$ and $\langle$Fe$\rangle$.  Dotted lines connect constant-age models
(from top to bottom: 1.2, 2.2, 3.5, 7.0, and 14.1 Gyr) while solid
lines connect models with the same [Fe/H] (from left to right: -1.3,
-0.7, -0.4, 0.0, and +0.2).  The square shows a sample data point for
a galaxy with age $\approx 7$ Gyrs and [Fe/H] $\approx -0.2$.  In this
figure, it is easy to see which box of the grid encloses the data
point in question.  This gives a bounded range in age and [Fe/H] for
the data point.  A two-dimensional linear interpolation within the
gridbox is then used to convert the data point in index-index space
into fiducial values for the age and [Fe/H].

\subsection{Determining Non-Solar Abundance Ratios}  \label{method}

\begin{figure}
\epsscale{1}
\plotone{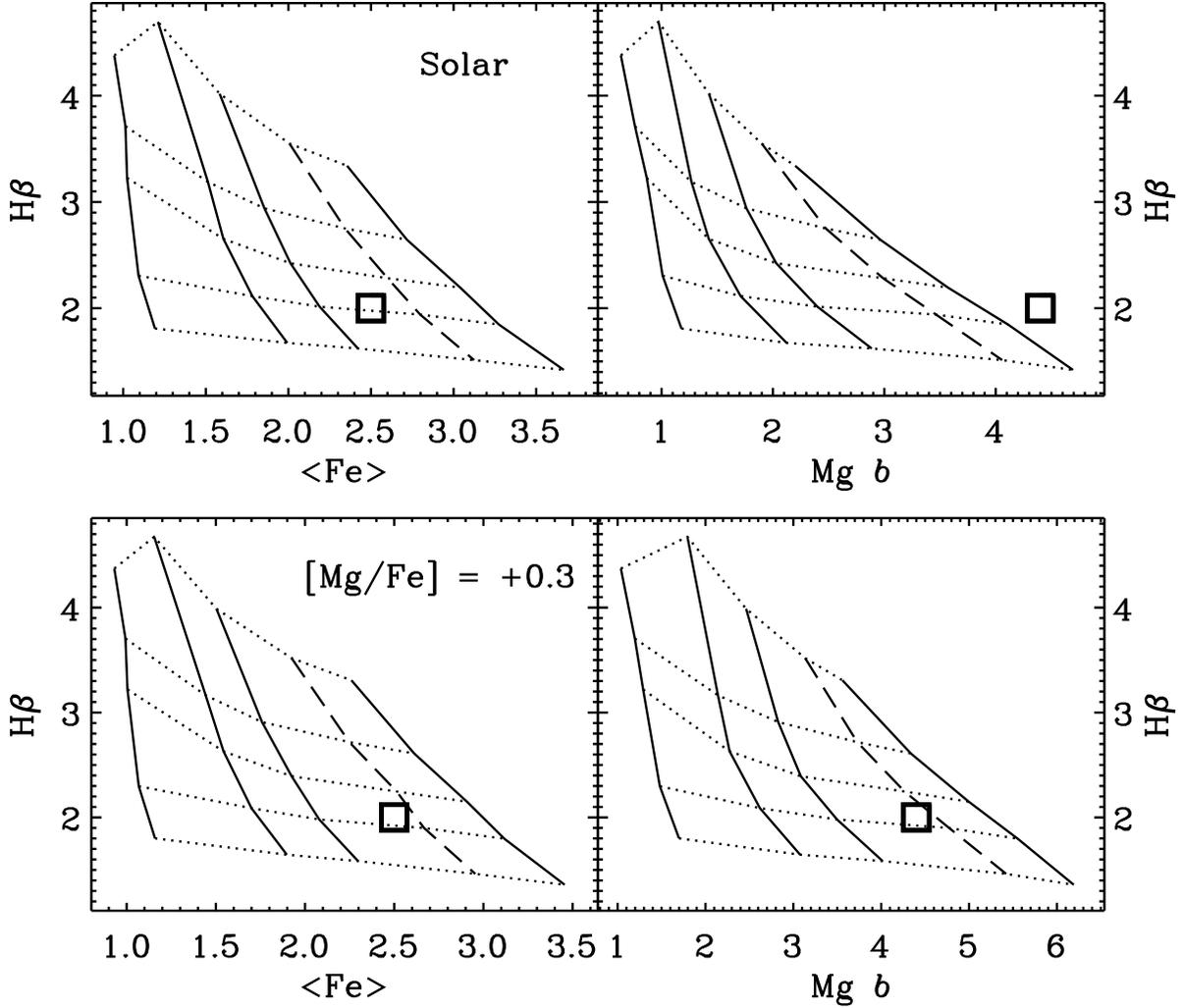}
\caption{Grids produced by the S07 model illustrating the
  effects of super-solar [Mg/Fe].  Solid lines show constant [Fe/H]
  from left to right of -1.3, -0.7, -0.4, 0.0, and +0.2 ([Fe/H] = 0.0
  is shown as a dashed line for reference).  Dotted lines show
  constant age from top to bottom of 1.2, 2.2, 3.5, 7.0, and 14.1 Gyr.
  The square shows an example data point and is the same in both
  panels.  {\it Upper panel:} Models computed with solar-scaled
  abundances.  The Balmer-Fe grid gives fiducial values of $t = 7$ Gyr
  and [Fe/H] = -0.2.  The Balmer-Mg~{\it b} grid shows an [Fe/H] $>
  0.4$ dex higher than the fiducial from the Balmer-Fe grid.  This
  indicates that [Mg/Fe] is super-solar.  {\it Lower panel:}
  Increasing [Mg/Fe] in the models slides the grid to the right toward
  the Balmer-Mg~{\it b} data point, lowering the estimated [Fe/H].
  H$\beta$ is also slightly affected by the abundance change, yielding
  a slightly younger fiducial age.  In the Mg-enhanced model, the
  fiducial [Fe/H] from the Balmer-Fe grid matches the value of [Fe/H]
  estimated from the Balmer-Mg~{\it b} grid, indicating that [Mg/Fe]
  $=+0.3$ is a good fit to the data. }\label{grids}
\end{figure}

Once a fiducial age and [Fe/H] have been determined from the
H$\beta$--$\langle$Fe$\rangle$ model grid, one can then make a similar grid plot
with an index which is sensitive to non-solar abundance ratios, for
instance a Balmer line plotted against Mg~{\it b}.  Mg~{\it b} is
strongly affected by the [Mg/Fe] abundance ratio, at fixed [Fe/H].  If
the chosen S07 model is a good match to the data, then the age and
[Fe/H] estimated by grid inversion from the H$\beta$--Mg~{\it b} plot
should match the fiducial age and [Fe/H] from the H$\beta$--$\langle$Fe$\rangle$
plot.  In our example in Figure \ref{grids}, this is clearly not the
case; the H$\beta$--Mg~{\it b} plot in the top right panel of Figure
\ref{grids} gives a larger value of [Fe/H] than the H$\beta$--$\langle$Fe$\rangle$
plot in the top left panel. This is an indication that the example
galaxy has a larger [Mg/Fe] ratio than is predicted in a solar-scaled
model.  The sequential grid inversion algorithm then increases the
input value of [Mg/Fe] and recomputes the S07 model.  Increasing
[Mg/Fe] will increase the strength of the Mg~{\it b} index {\it at
fixed [Fe/H]}, which will have the effect of sliding the entire set of
model grids to the right in the H$\beta$--Mg~{\it b} plot, while
keeping them essentially unchanged in the H$\beta$--$\langle$Fe$\rangle$ plane.
This~{\it lowers} the value of [Fe/H] that is estimated by grid
inversion in the H$\beta$--Mg~{\it b} plot, bringing it into agreement
with that obtained in the H$\beta$--$\langle$Fe$\rangle$ diagram.  This can been
seen in the lower panels of Figure \ref{grids}, which show
H$\beta$--$\langle$Fe$\rangle$ and H$\beta$--Mg~$b$ plots for an S07 model
computed with [Mg/Fe]$=+0.3$.  For the data shown in this plot, the
age and [Fe/H] values indicated by the model grids are consistent for
both metal lines.  In the sequential grid-inversion algorithm
implemented in EZ\_Ages, [Mg/Fe] is increased incrementally until the
[Fe/H] estimated from the H$\beta$--Mg~{\it b} plot matches the
fiducial [Fe/H] from the H$\beta$--$\langle$Fe$\rangle$ plot.  This process is
then repeated for other Lick indices and other abundance ratios.

The key to the sequential grid inversion algorithm is to proceed with
the abundance fitting in such a way as to only adjust one abundance at
a time.  Once fiducial values for the age and [Fe/H] of the system
have been determined, the goal is to next fit a Lick index that
introduces only one additional abundance.  Mg~{\it b} is a good choice
because it is dominated by Mg and Fe (see Table 1), with the latter
being determined from the H$\beta$--$\langle$Fe$\rangle$ grid
inversion.  By adjusting [Mg/Fe] until the H$\beta$--Mg~{\it b} grid
inversion matches the fiducial values, EZ\_Ages can find the
best-fitting Mg abundance.

Another index which EZ\_Ages can use to fit a single element is
C$_2$4668, which is dominated by C and O.  Oxygen is notoriously
difficult to estimate in old (ie. non-star forming) unresolved stellar
populations, therefore EZ\_Ages does not try to model [O/Fe] (but see
discussion in \S\ref{consist}).  Instead, it allows the user to choose
an input oxygen abundance.  For consistency, this should be chosen to
match the input isochrone: [O/Fe] = 0.0 for the solar-scaled
isochrone, [O/Fe] = +0.5 for the $\alpha$-enhanced isochrone.  With
[O/Fe] set by the user, this leaves only C as a variable abundance
which contibutes strongly to C$_2$4668.  EZ\_Ages adjusts [C/Fe] so
that the [Fe/H] estimated in the H$\beta$--C$_2$4668 plot
matches the fiducial.

The only indices which are strongly affected by nitrogen and which
might therefore allow a fit of [N/Fe] are the two CN lines, CN$_1$ and
CN$_2$, both of which respond strongly to changes in C, N, and O.  As
discussed above, [O/Fe] is set by the user.  This leaves C and N as
variable abundances contributing to the CN indices.  However, if
[C/Fe] has already been determined by fitting C$_2$4668, then the CN
indices can be used to fit for [N/Fe] at the calculated value of
[C/Fe].  The computed value of [N/Fe] is therefore dependent upon the
calculated value of [C/Fe] and a robust error analysis should propagate
errors from [C/Fe] into the the errors in [N/Fe].  

The S07 models include one Lick index dominated by Ca and can
therefore be used to fit [Ca/Fe]: Ca4227.  Because the blue continuum
region used in measuring Ca4227 overlaps with a CN absorption band,
this index is also very sensitive to [C/Fe] and [N/Fe], so that these
abundance ratios must be fit properly before using Ca4227 to determine
[Ca/Fe].  As with [N/Fe], errors in the calculated values of [C/Fe]
and [N/Fe] contribute to the error in [Ca/Fe].  Again, this should be
taken into consideration in a complete error analysis (see
\S\ref{errors}).

The set of available Lick indices and their element abundance
sensitivities thus prescribe an order for abundance fitting: [C/Fe],
then [N/Fe], then [Ca/Fe].  Note that Mg~{\it b}, the preferred
index used to fit [Mg/Fe], does not depend strongly on these
abundances, nor do the other abundances depend on it, therefore it
can be fit in any order.

Based on Table \ref{indextable}, it is clear that some indices are
preferable for fitting: those that are dominated by the elemental
abundances that we are using them to estimate and no others.  Because
of this, EZ\_Ages has a built-in hierarchy which it uses to choose the
indices used in abundance fitting.  Among the Balmer lines, the
preferred index is H$\beta$ because it is mildly sensitive to total
metallicity, but relatively insensitive to individual abundances.  It
is followed by H$\delta_F$ and H$\gamma_F$, which are somewhat
sensitive to individual elements, and lastly H$\delta_A$ and
H$\gamma_A$, which are broader indices (designed to measure Balmer
strengths in populations with a substantial A star component) and
therefore include a larger number of other absorption features in the
index and pseudo-continuum bandpasses.  

For iron indices, the order of preference is $\langle$Fe$\rangle$ (an
average of Fe5270 and Fe5335), Fe5270, Fe5335, Fe4383, and Fe5015.
The indices Fe5270 and Fe5335 are dominated by Fe and total
metallicity effects, making them relatively clean indicators of
[Fe/H].  Of the other two lines, both of which are influenced by
element abundance ratios, we prefer Fe4383 over Fe5015 for two
reasons: Fe5015 falls on a CCD defect in the globular cluster data
used in \S\ref{compare_clusters} to test the output of EZ\_Ages and
thus can only be tested for M~67, where it appears to perform
similarly well to Fe4383, and the Fe5015 bandpass overlaps the
[O\textsc{iii}]$\lambda$5007 emission line from ionized gas sometimes
present in galaxies, thus limiting its usefulness in stellar
population modeling of galaxies.

For fitting [Mg/Fe], the preferred index is Mg~{\it b} followed by
Mg$_2$, which is more sensitive to C than is Mg~{\it b}.  For [C/Fe],
it is C$_2$4668 followed by G4300, for reasons discussed in detail in
\S\ref{consist}.  For [N/Fe], it is CN$_1$ followed by CN$_2$,
although this is fairly arbitrary as there is no strong reason to
prefer one CN index over the other.  For [Ca/Fe], the only available
index is Ca4227.

It is possible to manually override this order of preference if a
given index is unavailable, or to investigate the effects of using
other lines in the fit.  Note that EZ\_Ages will only fit an abundance
ratio if at least one of the indices needed to estimate it is
available.  Also, abundances which depend on other abundance ratios
will not be fit if the prerequisite abundance is not available (e.g.,
EZ\_Ages will not use a CN index to fit [N/Fe] unless [C/Fe] has been
determined from C$_2$4668 or G4300).

Thus far, we have focused only on the dominant abundances affecting
each index, however many indices have a weak dependence on other
abundances.  Particularly if the most desirable Lick indices are not
available due to limited wavelength coverage, bad skyline subtraction,
or nebular emission within the galaxy and other lines have to be used
(such as H$\delta_F$ in place of H$\beta$ or Fe4383 in place of Fe5270
or Fe5335), the computed fiducial values for age and [Fe/H] or other
abundance values may change once the abundance fitting is complete,
because some of the newly determined abundances may affect the indices
used to determine those fiducial values.  To ensure that the final
abundance ratio is consistent for all of the Lick indices used in the
fit, EZ\_Ages iterates the fitting process up to four times.  It uses
the best-fitting abundance pattern from the previous iteration to
compute new fiducial values for age and [Fe/H], then repeats the
fitting process until further iterations do not improve the fit.

The S07 models account for variations in Lick index strength due to
abundance variations in the elements Fe, Mg, C, N, Ca, O, Na, Si, Cr,
and Ti.  However, only the first five of these elements can be fit
using the algorithm described above.  The elements O, Na, Si, Cr, and
Ti do not dominate any of the Lick indices modeled by S07 and
therefore cannot be fit in this process.  The EZ\_Ages code allows the
user to supply a value for any of these unmodeled elements to use as
input to the S07 models.  In the absence of user-supplied values,
EZ\_Ages has a set of default abundances for these elements.  By
default, it sets [O/Fe] to match the chosen isochrone ([O/Fe] = 0.0
for the solar-scaled isochrones, [O/Fe] = $+0.5$ for the
$\alpha$-enhanced isochrone).  The default for the $\alpha$-elements
Na, Si, and Ti is that they are set to follow Mg, so that [Na/Fe] =
[Si/Fe] = [Ti/Fe] = [Mg/Fe].  The iron-peak element Cr is set to
follow Fe, so that [Cr/Fe] = 0.0.  All of the modeling reported in
this analysis was done using the default settings for the unmodeled
elements and with the solar-scaled isochrone.  To see the effect of
using the $\alpha$-enhanced isochrone or super-solar [O/Fe] on
abundance fitting, see \S4.3 of S07 and Figures 12 and 14 in
\citet{gra07}.

\subsection{Error Calculations} \label{errors}

If errors in the Lick index data are provided as input, EZ\_Ages
uses them to estimate the uncertainties in the ages, [Fe/H], and
abundance ratios calculated by the sequential grid inversion
algorithm.  Uncertainties are assumed to be dominated by measurement
errors in the line strengths.  Systematic uncertainties in the models
are ignored, except that the usual caveat applies: comparisons of {\it
relative} ages and abundances between multiple objects are more
reliable than absolute age or abundance estimates.

Uncertainties in age and [Fe/H] due to line strength measurement
errors are computed by displacing the measured data point by the input
index error, then redoing the grid inversion on the displaced data
point to determine the range in age or [Fe/H] constrained by 1$\sigma$
error in the measured Balmer or Fe line strength.  Because the grid
spacing is non-linear (particularly in age), errors in the positive
and negative direction are computed separately.  It has been noted by
many previous authors that the errors in SSP age and [Fe/H] as
inferred from index-index plots are not independent due to the fact
that the model grids are not orthogonal.  As can be seen from an
inspection of Figure \ref{grids}, age and [Fe/H] errors are correlated
in the sense that lower inferred ages will lead to higher inferred
[Fe/H], and vice versa.  EZ\_Ages does not explicitly model this
aspect of the inferred errors in age and [Fe/H] values, but users
should keep this effect in mind when interpreting results from SSP
index-index fitting.

Any error in the choice of fiducial age and [Fe/H] will result in
errors in the abundance ratios estimated by fitting to incorrect
values of the age or [Fe/H].  Thus uncertainties in the fiducial age
and [Fe/H] must be propagated through to determine their effect on the
abundance ratio estimates.

For example, EZ\_Ages assumes that the uncertainty in [Mg/Fe] is
dominated by two sources: the measurement error of the Mg-sensitive
Lick index used in the abundance ratio determination (e.g., Mg~{\it
b}) and the uncertainty in the fiducial [Fe/H] which EZ\_Ages
attempts to match in the H$\beta$--Mg~{\it b} plot.  In the following
discussion, we will use $\Delta$Mg~{\it b} and $\Delta$[Fe/H] to
represent the measurement error in Mg~{\it b} and the uncertainty in
the fiducial [Fe/H], respectively.  The uncertainty in [Mg/Fe] due to
Mg~{\it b} measurement error and fiducial uncertainty are denoted in
turn by $\Delta$[Mg/Fe]$_{\rm{Mg}b}$ and
$\Delta$[Mg/Fe]$_{\rm{fid}}$.

To determine $\Delta$[Mg/Fe]$_{\rm{Mg}b}$, EZ\_Ages recomputes
the [Mg/Fe] abundance for Mg~{\it b}$+\Delta$Mg~{\it b} and Mg~{\it
b}$-\Delta$Mg~{\it b}, using the larger of these two uncertainties as
the estimated $\Delta$[Mg/Fe]$_{\rm{Mg}b}$.  To determine
$\Delta$[Mg/Fe]$_{\rm{fid}}$, EZ\_Ages recomputes [Mg/Fe] using
the measured value of Mg~{\it b} to fit [Fe/H]$+\Delta$[Fe/H] and
[Fe/H]$-\Delta$[Fe/H], again choosing the larger uncertainty for
estimating $\Delta$[Mg/Fe]$_{\rm{fid}}$.  These two sources of error
are treated as independent and the total $\Delta$[Mg/Fe] is determined
by summing in quadrature $\Delta$[Mg/Fe]$_{\rm{Mg}b}$ and
$\Delta$[Mg/Fe]$_{\rm{fid}}$.  

The process for computing $\Delta$[C/Fe] is identical to that for
[Mg/Fe], substituting a C-sensitive Lick index for Mg~{\it b}.
Because [N/Fe] is determined using a CN-sensitive index such as
CN$_1$, $\Delta$[N/Fe] depends not only on the CN$_1$ line strength
measurement error and $\Delta$[Fe/H], but on $\Delta$[C/Fe] as well.
The contributions to $\Delta$[N/Fe] due to $\Delta$CN$_1$ and
$\Delta$[Fe/H] (i.e., $\Delta$[N/Fe]$_{\rm{CN1}}$ and
$\Delta$[N/Fe]$_{\rm{fid}}$) are computed as above.  An additional
contribution due to $\Delta$[C/Fe] is determined by matching the
measured CN$_1$ line strength to the fiducial [Fe/H], using models
with carbon abundances given by [C/Fe]$+\Delta$[C/Fe] and
[C/Fe]$-\Delta$[C/Fe].  These three sources of error are summed in
quadrature to obtain $\Delta$[N/Fe].  A similar process is repeated
for [Ca/Fe], with the sole difference that errors in [N/Fe] also
contribute to the [Ca/Fe] error budget.

To check that this analysis provides a reasonable estimate of the
actual statistical errors, we have used a Monte Carlo method to
generate a large set of artificial ``repeat measurements'' of the Lick
indices for a test stellar population and compared the resulting
distribution of ages, abundances, and abundance ratios to the errors
calculated by EZ\_Ages.  For our test data, we used Lick index values
measured in a stack of $\sim$10,000 red sequence galaxy spectra taken
from the Sloan Digital Sky Survey.  As modeled by {\it EZ\_Ages},
these correspond to a stellar population with a mean light-weighted
age of 9.1 Gyr, with [Fe/H]$=-0.15$, [Mg/Fe]$=+0.17$, [C/Fe]$=+0.17$,
[N/Fe]$=+0.07$, and [Ca/Fe]$=0.00$.

We ran four Monte Carlo simulations for different error ranges
(effectively different $S/N$ spectra), in which we assumed the errors
were 2\%, 5\%, 10\%, and 20\% of the measured index values.  For each
run, we simulated 100 repeat measurements of the data by choosing
values for each Lick index drawn from Gaussian distributions centered
on the ``true'' data values with widths characterised by the assumed
2\%, 5\%, 10\%, and 20\% errors.  This produced a simulation of 100
realizations for each of the four assumed error ranges.  {\it
EZ\_Ages} was used to determine the age, [Fe/H], and abundance ratios
for each simulated measurement.  The distribution of these values for
the 100 simulated repeat measurements can be compared with the error
estimate made by EZ\_Ages for the original data values and index
measurement errors.  

\begin{figure}
\epsscale{1.0}
\plotone{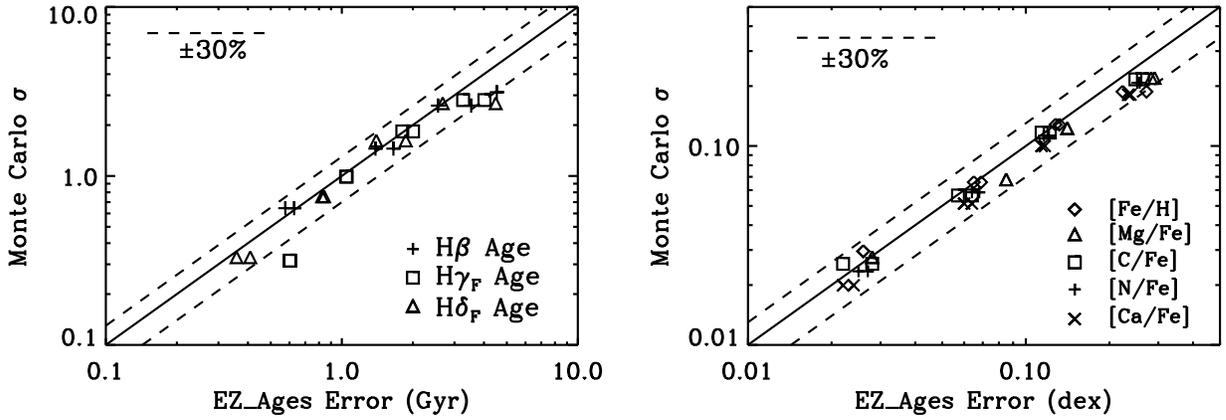}
\caption{ Comparison of the error estimates determined by EZ\_Ages
  with the results of the Monte Carlo error simulation.  Age errors
  are plotted in the left panel and [Fe/H] and abundance errors in the
  right panel.  The solid lines show one-to-one identity relations,
  with a $\pm30$\% spread indicated by dashed lines.  The error
  estimates produced by EZ\_Ages are consistent with those from the
  full Monte Carlo simulations, indicating that EZ\_Ages produces
  reliable error estimates.
}\label{errplot}
\end{figure}

This comparison is shown in Figure \ref{errplot}.  The errors reported
by EZ\_Ages are shown on the x-axis, while the y-axis shows the
standard deviation $\sigma$ of the distributions of ages and
abundances produced in each Monte Carlo simulation.  As discussed
earlier in this section, EZ\_Ages calculates separately errors in the
positive and negative direction.  There are thus two error values
produced by EZ\_Ages plotted against each Monte Carlo $\sigma$.  The
solid line shows a one-to-one relation.  The dashed lines outline
error estimates that are within $\pm30$\% of the 1$\sigma$ Monte Carlo
spread.

For the simulations with 2\%, 5\%, and 10\% index measurement errors,
the age and abundance ratio error estimates from EZ\_Ages are all
within 30\% of the results of the Monte Carlo simulation with the
exception of the error in the H$\gamma_F$ age, which is overestimated
by about 0.3 Gyr in the 2\% error case.  Although this is a
substantial fraction of the simulated H$\gamma_F$ age error, it is
well within the uncertainties in the modeling process, which are of
order 1 Gyr in the $\sim9$ Gyr age range corresponding to the test
data point \citep{sch07}.  In all other cases, the EZ\_Ages error
estimate is an excellent match to the 1$\sigma$ errors produced by the
full Monte Carlo realization.  For the simulations with 20\% index
measurement errors, EZ\_Ages tends to slightly overestimate the errors
as compared to the Monte Carlo simulation, but in all cases is still
within 35\% of the simulated error.

The good match shown in this test between the error estimates
generated by EZ\_Ages and those produced by a Monte Carlo simulation
of the errors suggests that EZ\_Ages does in fact do a reasonable job
of estimating the errors in the age, [Fe/H], and abundance ratio
values calculated.  This includes the propagation of errors between
various stages of the modeling process.  As the data become noisier,
EZ\_Ages has a tendency to slightly overestimate the actual
uncertainty in the age, [Fe/H], and abundance due to measurement
errors.  Figure \ref{errplot} also demonstrates that in order to
estimate ages, [Fe/H], and abundance ratios within $\pm$0.1 dex for a
typical early type galaxy, Lick indices must be measured with $\pm5$\%
accuracy, which requires spectra with $S/N \sim 100$ {\AA}$^{-1}$
\citep{car98}.

\subsection{Correlation of Errors in Derived
  Parameters}\label{correlated_errs} 

The Monte Carlo error simulation also allows us to investigate the
effect of correlated errors in the EZ\_Ages analysis.  Because lines
of constant age and [Fe/H] are not orthogonal in
H$\beta$-$\langle$Fe$\rangle$ space, errors in either index
measurement will result in anti-correlated errors in the inferred age
and [Fe/H], as discussed in detail in \citet{tra00}.  For example, an
error in H$\beta$ that causes the determined SSP age to be too old
will cause the measured [Fe/H] value to be too low, as can be seen
from the grids in Figure \ref{grids}.  Likewise, the sequential grid
inversion process for fitting abundance ratios depends on the fiducial
age and particularly on the fiducial [Fe/H] determination.  This will
tend to produce an anti-correlation between [Fe/H] and the measured
abundance ratios.

As shown in Figure \ref{errplot}, the simplified error propagation
implemented in EZ\_Ages does an excellent job of tracking the total
error in each stellar population parameter throughout the fitting
process.  However, it does not produce a full error ellipse
illustrating correlated errors.

\begin{figure}
\epsscale{1.0}
\plotone{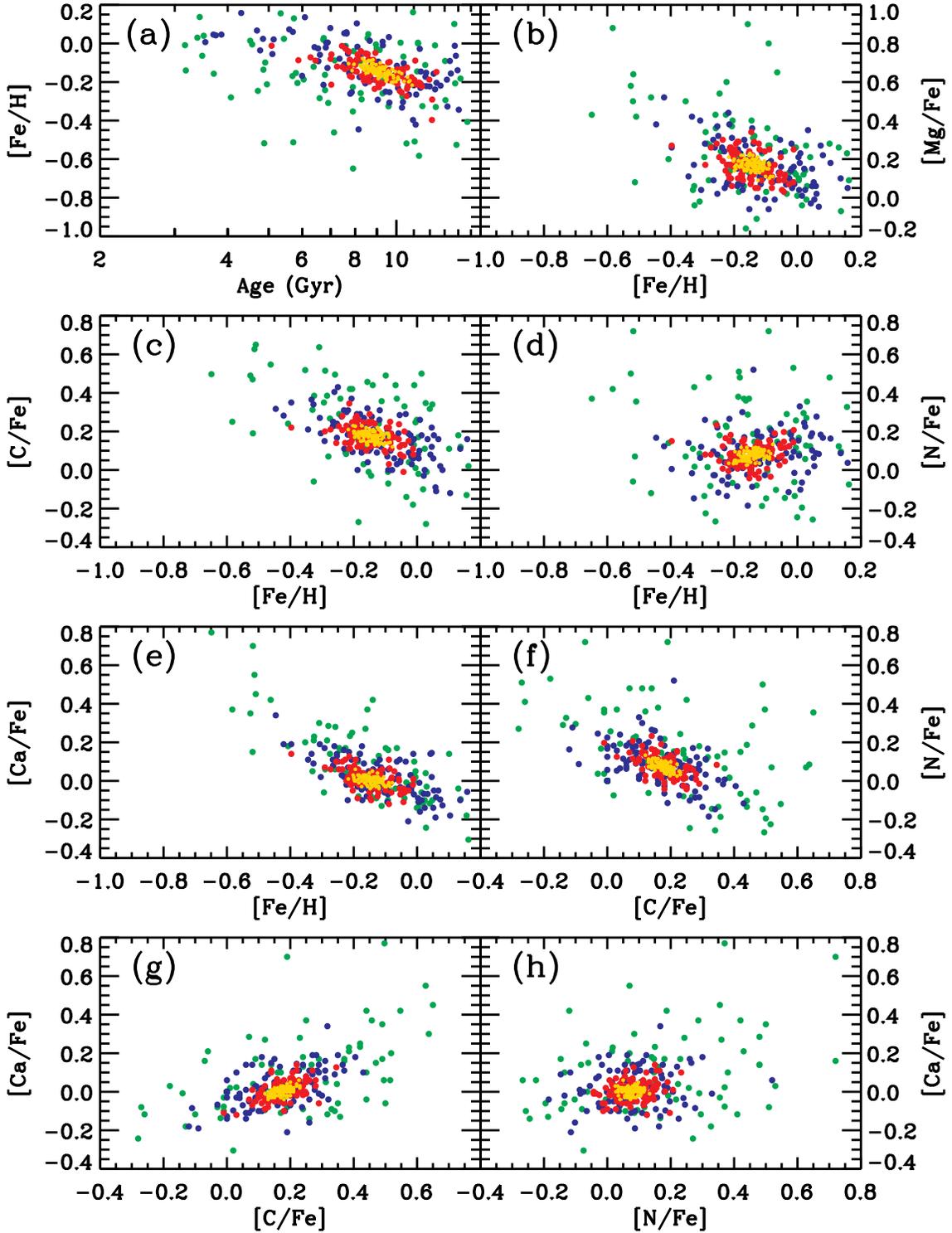}
\caption{The distribution of EZ\_Ages results for the Monte-Carlo
  error simulation.  Yellow, red, blue, and green points correspond to
  the realizations of a single data point drawn from the 2\%, 5\%,
  10\%, and 20\% error simulations, respectively.  These trace out the
  error ellipses of the EZ\_Ages fitting analysis and reveal the
  effect of correlated errors. }\label{corr_errs}
\end{figure}

Figure \ref{corr_errs} uses the results of the Monte Carlo error
simulation to illustrate the correlated errors that result from
sequential abundance fitting.  The 100 simulated data realizations for
each of the assumed 2\%, 5\%, 10\%, and 20\% index measurement errors
(yellow, red, blue, and green points, respectively) effectively trace
out the error ellipses in stellar population parameter space.  Figure
\ref{corr_errs}a shows the well-known error anti-correlation between
the age and [Fe/H] measurements, as expected from the
non-orthogonality of the age and [Fe/H] model grids in
H$\beta$-$\langle$Fe$\rangle$ space.

Panels b--e show the abundance ratio determinations for [Mg/Fe],
[C/Fe], [N/Fe], and [Ca/Fe] against [Fe/H].  The error ellipses for
these data are expected to show an anti-correlation of errors because
an over-estimate of [Fe/H] will tend to result in an under-estimate of
the other abundances.  For example, if an error in the
$\langle$Fe$\rangle$ index causes the fiducial [Fe/H] determination to
be too high, the measured abundance ratios for other elements will
tend to be too low (at a higher [Fe/H], a lower value of [Mg/Fe] is
needed to match the same Mg~$b$ index).  This anti-correlation is
observed between [Fe/H] and the abundances [Mg/Fe], [C/Fe], and
[Ca/Fe].  Interestingly, there appears to be no correlation in the
errors in [Fe/H] and [N/Fe].  Because [N/Fe] is fit using a CN index,
it depends strongly on the value of [C/Fe] previously determined by
fitting either C$_2$4668 or G4300 and thus we expect a
anti-correlation of errors between [C/Fe] and [N/Fe].  This is indeed
the case, as illustrated in Figure \ref{corr_errs}f, and may explain
the lack of correlation in the errors in [N/Fe] and [Fe/H].  As in the
case of [N/Fe], the fitted values of [Ca/Fe] depend on other abundance
fits and thus may produce correlated errors.  These are illustrated in
Figure \ref{corr_errs}g,h.

The effect of correlated errors can produce spurious trends in a
dataset, thus the correlations illustrated here should be kept in mind
when interpreting abundance patterns from EZ\_Ages or a similar
abundance fitting process.  Of course, the best guard against these
effects is to perform abundance fitting only on spectra with $S/N \ge
100$ \AA$^{-1}$, corresponding to 5\% index errors (red points in
Figure \ref{corr_errs}) or better.

\section{Testing the Models on Cluster Spectra}\label{compare_clusters}

The validation of stellar population synthesis models depends
crucially on their ability to match the integrated properties of
simple systems, such as stellar clusters, which can be reasonably
approximated by single stellar populations.  In fact, this simple
truism disguises the considerable complexity of the task, as attested
by the vast literature dedicated to using cluster observations to
constrain stellar population synthesis models (S07; \citealt{lee05};
\citealt{pro04}; \citealt{sch04a,sch04b}; \citealt{mar03};
\citealt{sch02a,sch02b}; \citealt{bea02}; \citealt{vaz01};
\citealt{gib99}; \citealt{sch99}; \citealt{bru97}; \citealt{ros94};
and references therein).  S07 presented a lengthy discussion of the
comparison of his models with integrated data for well-known Galactic
open and globular clusters.  Models were computed for a range of ages
and metallicities, adopting the known abundance patterns of M~5,
47~Tuc, NGC~6528, and M~67. These were then compared with Lick index
measurements taken in the (very high S/N) cluster spectra.  As a
result, it was found that predictions for Balmer and metal lines
agreed with the data to within 1-$\sigma$ index measurement errors.
Also very importantly, the best-matching ages differed by no more than
$\sim$1~Gyr from the value obtained from analysis of cluster
color-magnitude diagrams (CMDs).  The only exception was M~5, for
which the age found was too low, due to the presence of blue
horizontal branch (HB) stars, which are not accounted for by the
models (see \S\ref{consist} for a further discussion of this point).

In this section we ask the inverse question: how well do the models
fare when used to convert a given set of index measurements into ages
and elemental abundances?  Again, we rely on data for clusters whose
abundances and ages are well constrained from analysis of individual
stellar spectra and deep cluster CMDs.  Two important model checks are
performed.  We start by submitting the models to the most challenging
of the two tests (\S\ref{real}), which consists of checking whether
the results obtained with EZ\_Ages match those from detailed resolved
analysis of high-resolution spectra and deep photometry of stellar
members.  While our {\it goal} is to match ages and the abundances of
Fe, Mg, C, N, and Ca to within 0.1 dex our {\it requirement} is that
differences be smaller than 0.2 dex.  Moreover, this set of
specifications should be met in a quick and non-interactive fashion,
suitable for applications to large data sets through use of EZ\_Ages.
As a second check (\S\ref{consist}), we verify the internal
consistency of EZ\_Ages, by comparing the results obtained for a given
set of observations, when using different combinations of Lick indices
as age and metal-abundance indicators.

The clusters of choice for this exercise are selected on the basis
of availability of good quality age and metal abundance determinations
(Table~\ref{abuntable}).  They span a wide range of metallicities,
from [Fe/H] $\sim$ --1.2 (NGC~6121) to nearly solar metallicity
(M~67 and NGC~6528), and include both old and intermediate-age
(M~67) clusters.  Lick indices for these clusters are presented in
Table \ref{cluster_indices}.

\subsection{Reality Check}  \label{real}

\begin{figure}
\epsscale{1.0}
\plotone{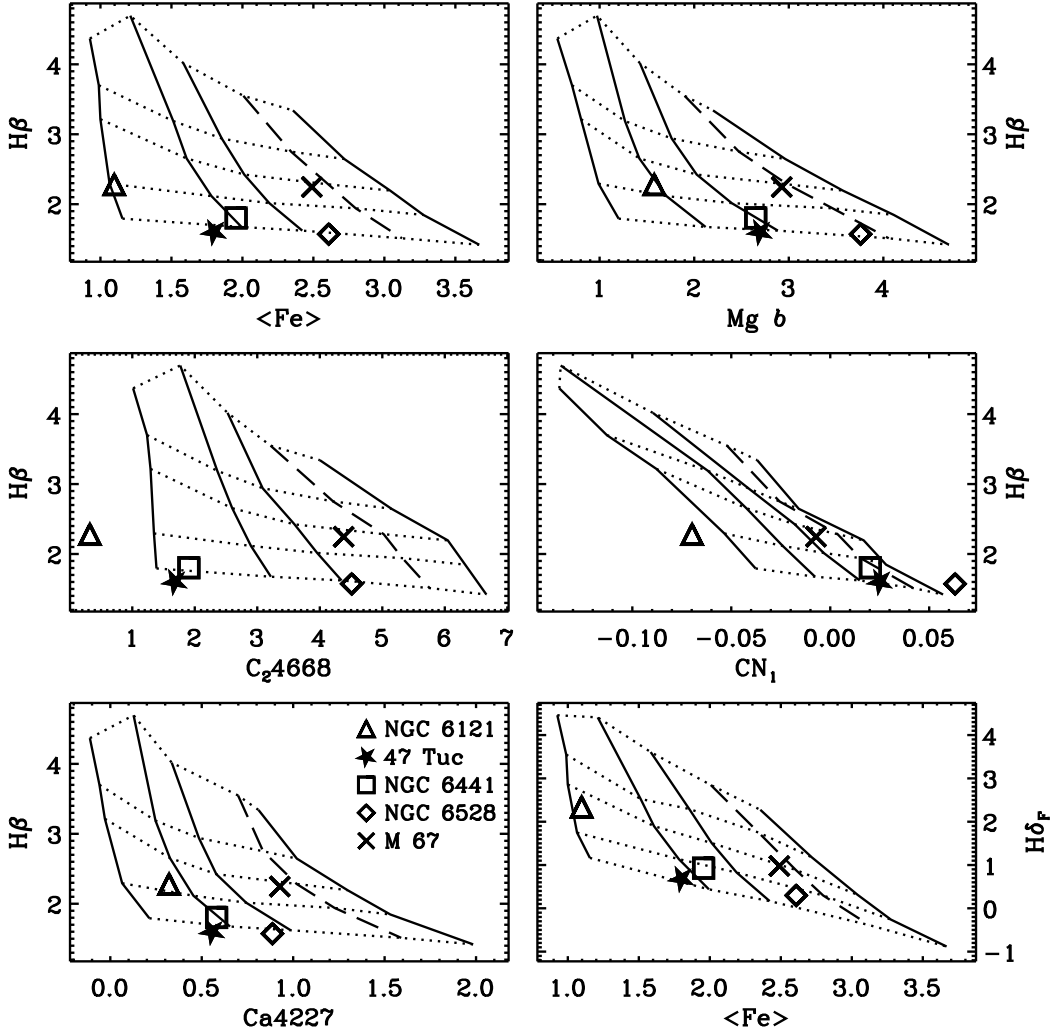}
\caption{S07 solar-scaled model grids compared with cluster data.
  Model grids lines are as in Figure \ref{grids}.  Because H$\beta$
  and $\langle$Fe$\rangle$ are relatively insensitive to non-solar
  abundance ratios, the fiducial age and [Fe/H] values for each
  cluster can be read off the model grid in the
  H$\beta$-$\langle$Fe$\rangle$ plot (upper left).  If solar-scaled
  models were a good fit to all clusters, the data would fall in the
  same region of the model space in each index-index plot, as is
  approximately the case with M~67.  The fact that other cluster data
  do not indicates that non-solar abundance patterns are needed to fit
  these clusters.  }\label{cluster_grids}
\end{figure}

In Figure~\ref{cluster_grids} cluster data are compared with the S07
solar abundance pattern models in a few index-index plots to
illustrate that solar-scaled models {\it cannot} match the full set of
observed cluster indices.  The indices displayed are the default age
and metal-abundance indicators employed by EZ\_Ages.  S07 showed that
H$\beta$, $\langle$Fe$\rangle$, Mg~$b$, C$_2$4668, CN$_1$ (or CN$_2$),
and Ca4227 are the most reliable, recommended indices for age, [Fe/H],
[Mg/Fe], [C/Fe], [N/Fe], and [Ca/Fe] determinations.  This will be
explored further in \S\ref{consist}.  Henceforth, we will refer to
this set of indices as the {\it standard set}.

The models displayed in Figure~\ref{cluster_grids} were computed for
the solar abundance pattern, and the data were measured in the
integrated spectra of the above globular clusters from the
\citet{sch05} spectral library and that of M~67, obtained by
\citet{sch04a}, and revised by S07.  In the upper left panel, model
and data are compared in the $\langle$Fe$\rangle$ vs. H$\beta$ plane.
S07 showed that these indices are excellent indicators of [Fe/H] and
age, respectively.  While the $\langle$Fe$\rangle$ average index is
relatively insensitive to the abundances of elements other than Fe,
H$\beta$ is mostly sensitive to age, being only mildly sensitive to
the overall metallicity of the stellar population.  Therefore, the age
and iron abundance of each cluster can be read fairly accurately from
the position of the cluster data on the model grid---except for the
case of NGC~6121 where the presence of blue HB stars affects the
H$\beta$-based age (see discussion in \S\ref{consist}).  The remaining
panels show plots of other metal-line indices against H$\beta$, and
$\langle$Fe$\rangle$ against H$\delta_F$.  In the metal-line vs.\
H$\beta$ plots, one can see that the positions on the model grids
where the data for most clusters fall are different from those they
occupy on the $\langle$Fe$\rangle$ vs. H$\beta$ plot.  As discussed in
\S\ref{method}, this is a clear indication that the abundance pattern
of the cluster stars is different from that of the Sun.  We apply the
method presented in the previous section to determine the abundance
pattern of the clusters.

\begin{figure}
\epsscale{1.0}
\plotone{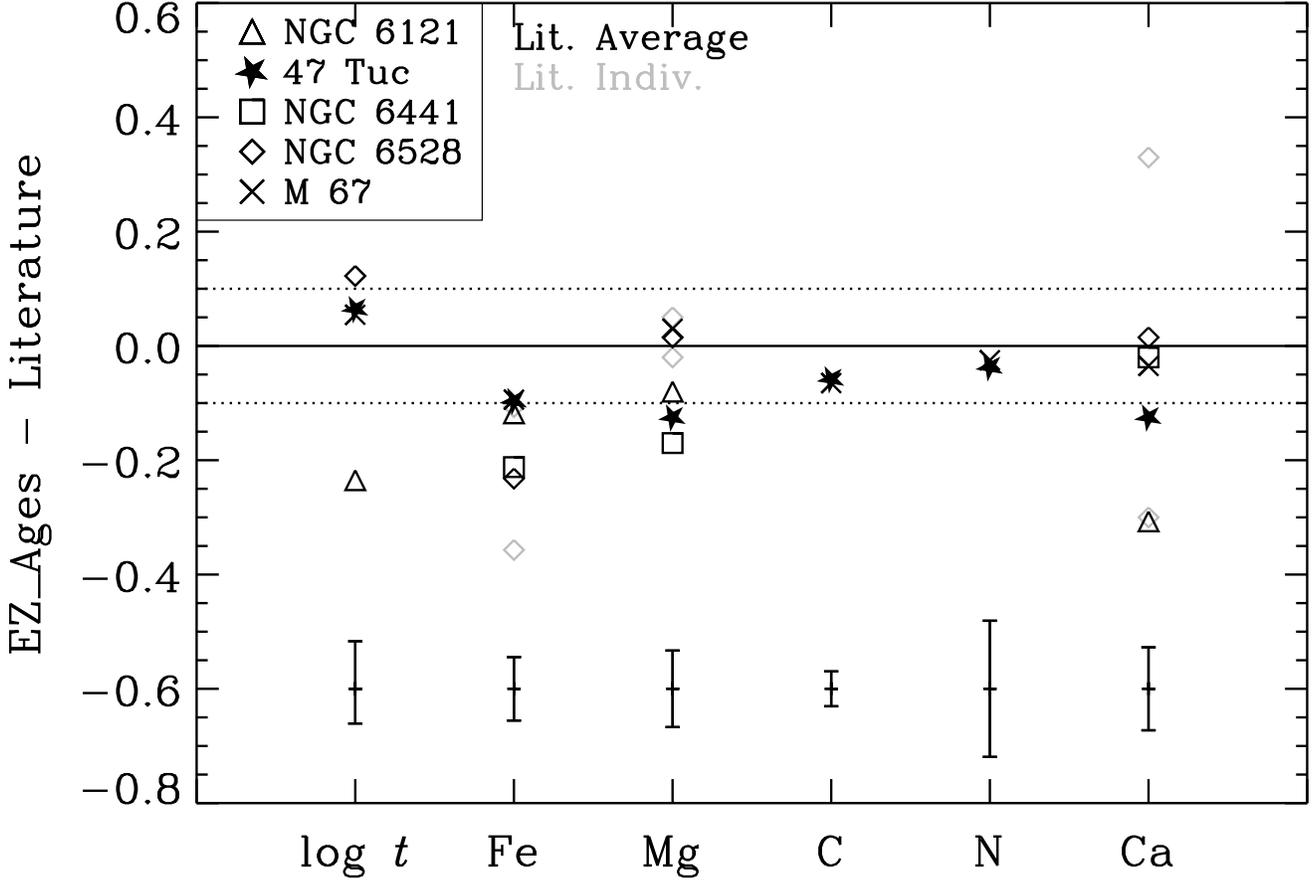}
\caption{Comparison of EZ\_Ages age ($\log t$), [Fe/H], [Mg/Fe],
  [C/Fe], [N/Fe], and [Ca/Fe] estimates with values from the
  literature.  Error bars indicate the mean observational errors for
  our cluster data.  Not all cluster parameters are available in the
  literature.  For NGC 6528, two sets of abundances are available from
  different groups and these differe substantially (see Table
  \ref{abuntable}).  The individual literature values are plotted in
  gray, with an average of the two literature values plotted in black.
  For 47~Tuc, [C/Fe] and [N/Fe] abundances in the literature are
  computed separately for CN-weak and CN-strong stars.  Here we
  compare the EZ\_Ages results for [C/Fe] and [N/Fe] against the
  average over both sets of stars, as this best simulates their
  combined contribution to the integrated light }\label{comp_lit}
\end{figure}

We ran EZ\_Ages on the cluster data presented in
Table~\ref{cluster_indices}. The results are listed in
Table~\ref{abuntable}, together with abundance and age determinations
from the literature.  The literature values are based on abundance
analyses of high-resolution spectra of individual cluster stars and on
deep cluster CMDs, respectively.  Differences between our results and
those from the literature are displayed in Figure~\ref{comp_lit}.  In
that plot, the dashed lines indicate the $\pm0.1$ dex goal stated in
the EZ\_Ages specifications discussed in \S\ref{compare_clusters}.
Average values from the literature are displayed in black and
individual determinations are shown in gray.  For carbon and nitrogen,
the values displayed are averages between CN-strong and CN-weak stars,
as this best approximates the effect of unresolved spectroscopic
observations.

A first inspection of the results shows that most abundance points
fall between the two dashed lines, indicating that we are meeting the
0.1-dex goal for most of the abundances and for most clusters.
Agreement is in fact particularly good for the abundances of carbon
and nitrogen, where the 0.1-dex goal was reached for all clusters for
which these elemental abundances are available in the literature.  For
magnesium and iron, we meet the 0.1-dex goal for some clusters, and
the 0.2-dex requirement for all of them, with one exception: for
NGC~6528, we find [Fe/H] = $-0.26$, in reasonably good agreement with
the value [Fe/H] = $-0.15$ reported by \citet{zoc04} and \citet{bar04}
but in significant disagreement with the value [Fe/H] = $+0.1$
reported by \citet{car01}.  For calcium, there is one cluster only for
which the 0.2-dex requirement is not met.  For all the others, we meet
the 0.1-dex goal.  Given that this is the first systematic attempt at
deriving abundances for all these five elements from integrated light,
these results can be considered very satisfactory.  Below we discuss
the few exceptions for which agreement is not so good.

Although the values of [Fe/H] determined by EZ\_Ages are almost all
within the 0.2-dex requirement, they are consistently slightly lower
than the literature abundances determined from invidual stars.  The
zeropoint for [Fe/H] in the models is based upon the [Fe/H]
measurements for the stellar library from which the S07 model is
constructed (the S07 models have not been separately calibrated to the
Solar spectrum) and the zeropoint is not known to perfect accuracy.  A
systematic 0.1-dex underprediction of [Fe/H] by the models is
possible.  However, this is not necessarily more uncertain than the
zeropoints in the cluster abundance determinations from individual
stars.  In the case of NGC~6528, for which multiple literature sources
are available, the abundance determinations in the literature show
substantial scatter: 0.25 dex in [Fe/H], 0.07 dex in [Mg/Fe], and 0.63
in [Ca/Fe].  In this context, it is difficult to address zeropoint
issues on the order of 0.1 dex by comparison with existing cluster
data.

In general we note that, for some clusters, elemental abundances
determined by different groups are in substantial disagreement, which
makes the task of interpreting our results a little more complicated.
That is especially the case of NGC~6528 mentioned above, for which
iron abundances from two groups differ by $\sim$ 0.25 dex, and calcium
abundances differ by more than 0.6 dex!  While a definitive statement
on our ability to match that cluster's abundances with EZ\_Ages awaits
the solution of this discrepancy, we call attention to the fact that
agreement with the average values is very good.

EZ\_Ages also performs very well in age determinations.  For all
clusters, with the sole exception of NGC~6121, the 0.2-dex requirement
is met, and for 47~Tuc and M~67, the 0.1-dex goal is also achieved.
For NGC~6121, the age obtained is too young because of the
contribution by blue HB stars which tend to boost the strength of
Balmer lines in the spectra of old stellar populations, mimicking a
younger age (see \S\ref{consist}).  The effect of blue HB stars on
Balmer lines has been pointed out in previous studies
(\citealt{fre95}; \citealt{mt00}; \citealt{sch04b}; \citealt{tra05};
\citealt{sch07}) and it can potentially lead to age underestimates.

The cluster NGC~6441 also contains blue HB stars \citep{ric97}, but
EZ\_Ages determines a relatively old age for the cluster when using
the H$\beta$ index in the modelling process, which seems to contradict
this interpretation.  Unfortunately, we could not find a CMD-based age
for this cluster in the literature so we cannot make a statement about
how the age result from EZ\_Ages compares to an independent age
estimate.  However, when the bluer Balmer line H$\delta_F$ is used in
the modelling of NGC~6441, EZ\_Ages finds a younger age than that
determined from H$\beta$, suggesting that the influence of the blue HB
stars does show up in H$\delta_F$.  As discussed by \citet{sch04b},
the signature of blue HB stars is stronger at bluer wavelengths, so
that they tend to impact age determinations based on H$\delta$ more
strongly than those based on H$\beta$.  This can be clearly seen in
the bottom right panel of Figure~\ref{cluster_grids}, where both
NGC~6121 and NGC~6441 look {\it younger} in the $\langle$Fe$\rangle$
vs.  H$\delta_F$ plot than in that involving H$\beta$.

In NGC~6441, the blue HB stars are greatly outnumbered by red HB
stars, unlike NGC~6121 in which the blue HB population is substantial.
They therefore have only a modest influence at the wavelength of
H$\beta$, but a larger influence at the bluer wavelength of
H$\delta_F$, where the blue HB stars contribute a greater fraction of
the total light \citep{sch04b}.  We further discuss this point in
\S\ref{consist}.

Finally, we note that the values we found for the abundances of C, N,
and Ca for NGC~6121 are suspiciously low.  Unfortunately, we could not
find literature values for C and N abundances to compare with our
estimates.  However, the numbers Table~\ref{abuntable} and
Figure~\ref{comp_lit} show that our [Ca/Fe] estimate is too low by
about 0.3 dex, which confirms our suspicions that there might be a
problem with our procedure for [Fe/H] $\simless -1.0$.  While this
clearly deserves further investigation, for the time being {\it we
strongly discourage users from applying EZ\_Ages in the [Fe/H]
$\simless -1.0$ regime}.

\begin{figure}
\epsscale{1.0}
\plotone{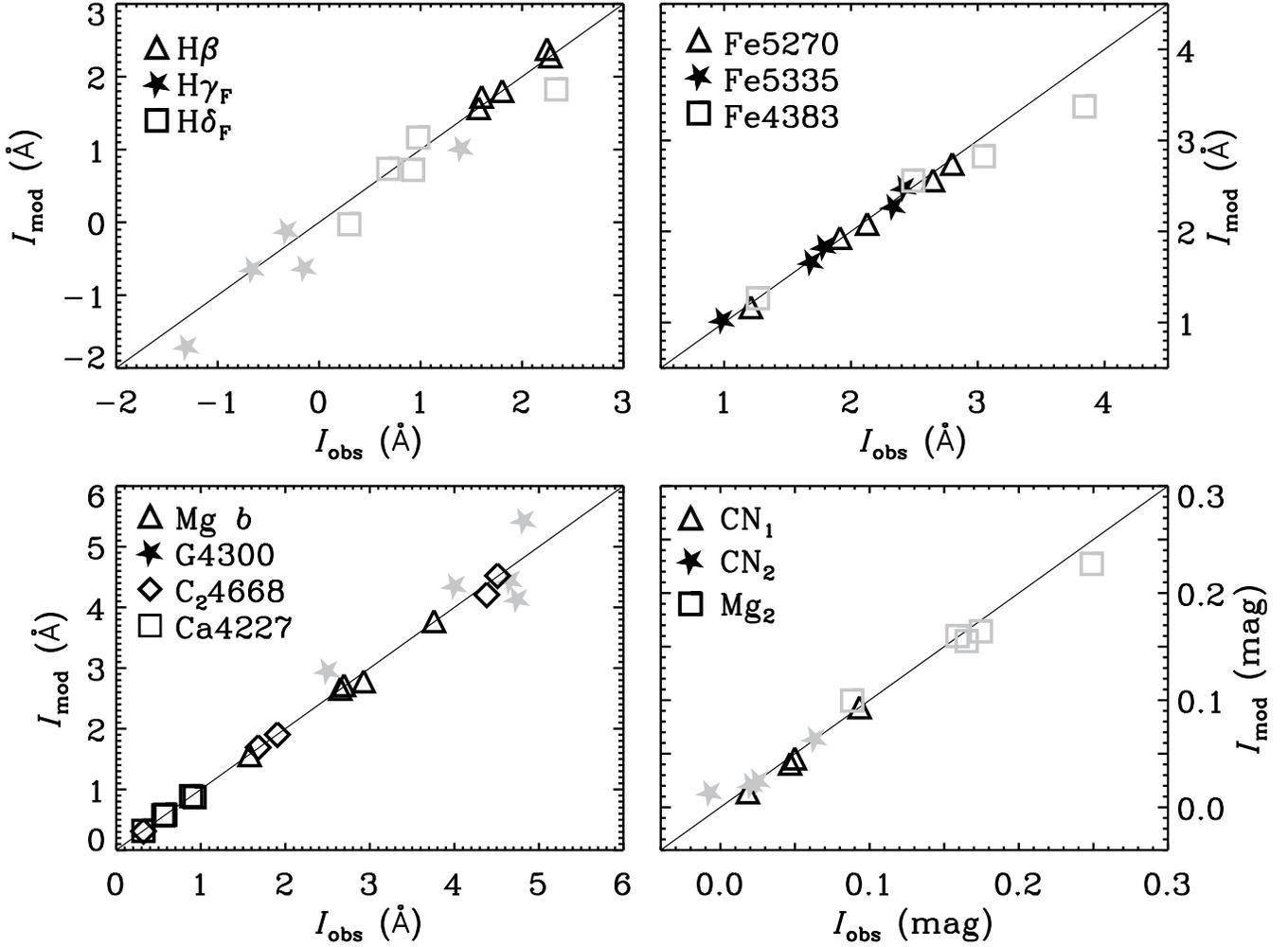}
\caption{Comparison of observed Lick index measurements with the
  predicted index values from the best-fitting S07 model, as
  determined by EZ\_Ages.  All indices included in the S07 models are
  shown for each of the five test clusters.  Measurement errors in the
  observed indices are approximately the size of the plotting symbols.
  Indices shown in black are those used in the abundance fitting
  process; those in gray are not used in the fitting.  Solid lines
  show the one-to-one relation.  Indices not used in the fitting
  process show some discrepancies between model predictions and the
  observed values, but there is no evidence for systematic
  discrepancies in the modeling of any of the
  indices. }\label{model_ind}
\end{figure}

Another check on the accuracy of the stellar population modeling is to
compare the index predictions of the best-fitting model to the
observed line strengths.  Figure \ref{model_ind} shows the observed
values of the Lick indices as measured in the five test cluster
spectra, plotted against the predicted index line strengths from the
best-fitting S07 model, as determined by EZ\_Ages.  Black and gray
symbols show indices that are and are not included in the fitting
process, respectively.  The indices used in the fitting process are
extremely well reproduced by the best-fitting model.  In addition, the
indices that are not used in the fitting process are also reasonably
well matched by the model predictions, with some scatter but with no
indication of systematic problems in the modeling of individual
indices.

\subsection{Consistency Check}  \label{consist}

\begin{figure}
\epsscale{1.0}
\plotone{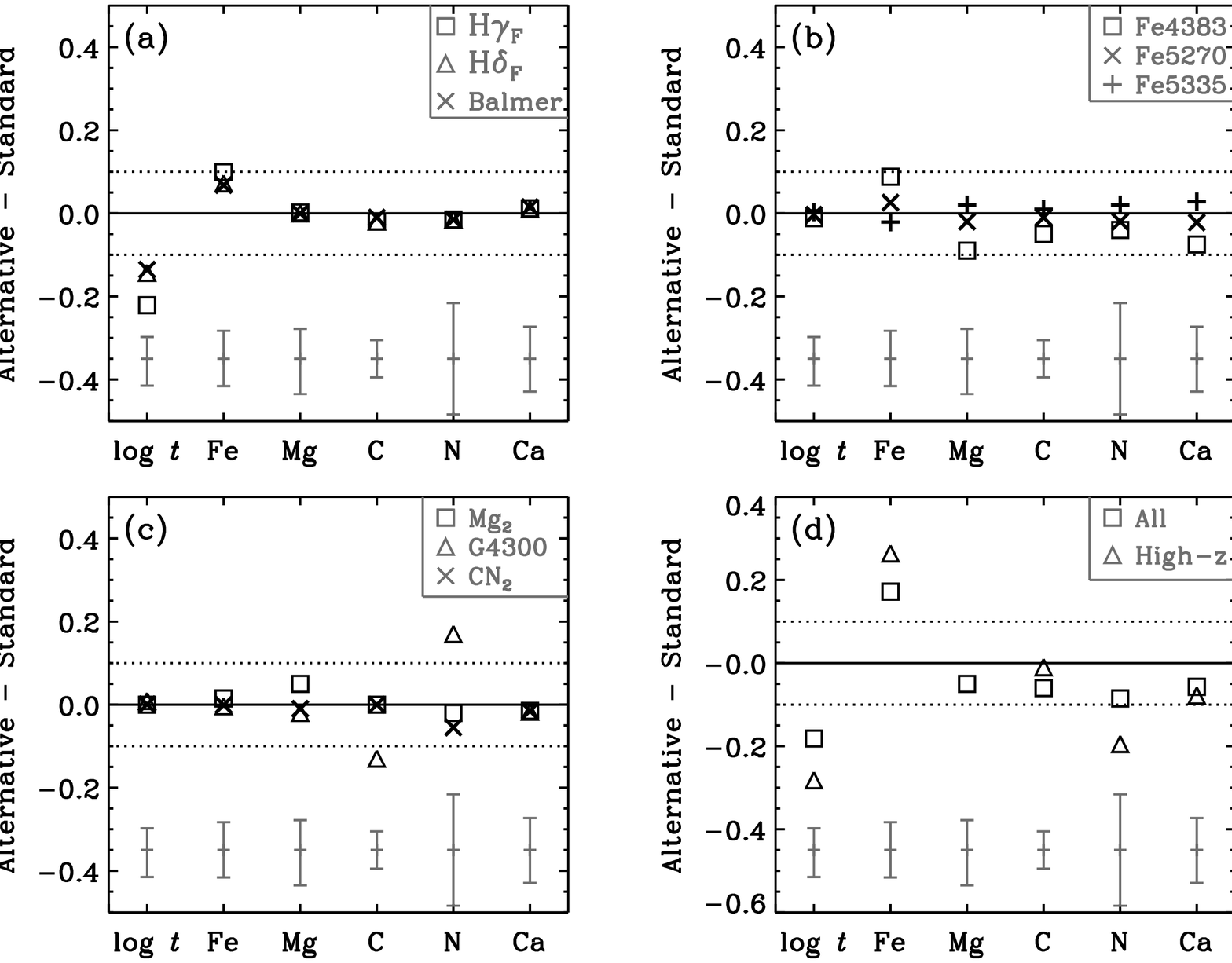}
\caption{Variations in age ($\log t$), [Fe/H], [Mg/Fe], [C/Fe],
  [N/Fe], and [Ca/Fe] estimates for NGC~6441 from using different
  indices in the fitting process.  Model names correspond to those in
  Table \ref{diff_ind_tab}.  Error bars indicate the
  observational errors calculated by EZ\_Ages for NGC~6441 using the
  standard set of indices.  (a--c) Results from substituting one
  alternative index into the standard index set (i.e., substituting
  G4300 for C$_2$4668 to fit [C/Fe]), and also for using averages of
  lines (i.e., an average of all Balmer lines instead of H$\beta$).
  All combinations of indices give results consistent within $\pm0.1$
  dex, with the exception of the bluer Balmer lines (panel a), which
  underpredict the cluster age, and G4300 (panel c, triangle), which
  underpredicts [C/Fe] by $-0.13$ dex relative to the fit using
  C$_2$4668.  The underprediction of [C/Fe] results in an
  overprediction of [N/Fe], as expected from the correlated errors
  between [C/Fe] and [N/Fe] (see Figure \ref{corr_errs}).  (d) Results
  for an average of all indices and for fitting when only indices
  blueward of 4400{\AA} are available (simulating higher redshift
  observations).  Fitting with Fe4383 and H$\delta_F$ instead of
  $\langle$Fe$\rangle$ and H$\beta$ (as in the simulated high redshift
  case) results in substantially younger ages and higher [Fe/H].  See
  text \S\ref{consist} for details.  }\label{vary_fit_NGC6441}
\end{figure}

\begin{figure}
\epsscale{1.0}
\plotone{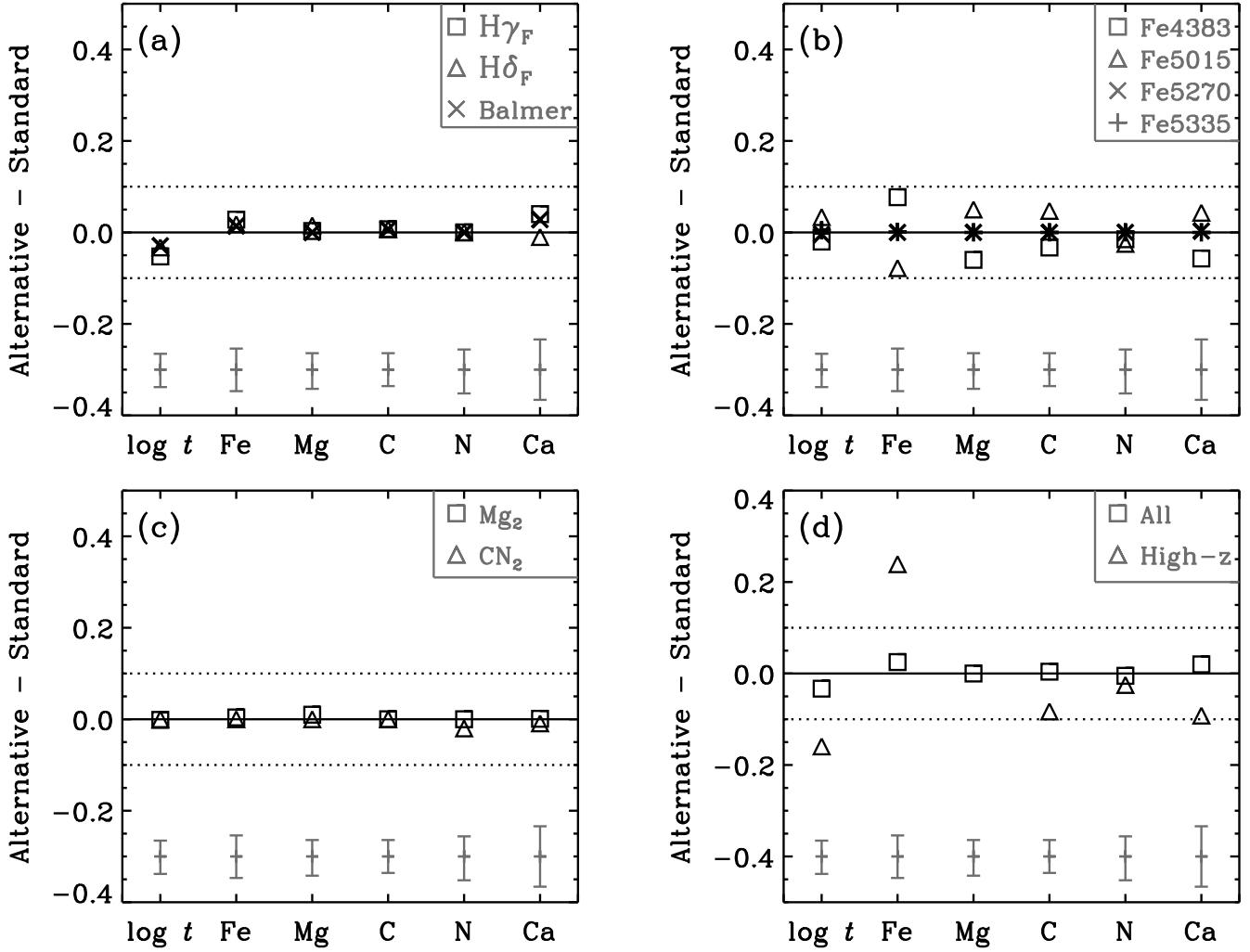}
\caption{Variations in age ($\log t$), [Fe/H], [Mg/Fe], [C/Fe],
  [N/Fe], and [Ca/Fe] estimates for M~67 from using different indices
  in the fitting process.  Model names correspond to those in
  Table \ref{diff_ind_tab}.  Error bars indicate the observational
  errors calculated by EZ\_Ages for M~67 using the standard set of
  indices.  (a--c) Results from substituting one alternative index
  into the standard index set (i.e., substituting G4300 for C$_2$4668
  to fit [C/Fe]), and also for using averages of lines (i.e., an
  average of all Balmer lines instead of H$\beta$).  (d) Results for
  an average of all indices and for fitting when only indices blueward
  of 4700{\AA} are available (simulating higher redshift
  observations).  Fitting with Fe4383 and H$\delta_F$ instead of
  $\langle$Fe$\rangle$ and H$\beta$ (as in the simulated high redshift
  case) results in substantially younger ages and higher [Fe/H].  See
  text \S\ref{consist} for details.  }\label{vary_fit_M67}
\end{figure}

In this test, we verify how the results vary when different index sets
are adopted as inputs to EZ\_Ages.  This is important, because the
full set of indices considered by EZ\_Ages are not always available to
observers, due to limitations such as those determined by the
instrumental setup adopted, or by the redshift of the sample studied.
Therefore, it is important to make sure that age and metal abundances
obtained from EZ\_Ages do not depend on which absorption-line indices
are employed---or, if they do, that the systematics is understood and
can be accounted for.  To test for this, we ran EZ\_Ages on multiple
combinations of line indices as measured for the clusters NGC~6441 and
M~67, and inter-compared the ages and metal abundances obtained.  As a
basis for the comparisons, we adopt the results obtained when using
the standard set, i.e., H$\beta$, $\langle$Fe$\rangle$, Mg~$b$,
C$_2$4668, CN$_1$, and Ca4227.  The indices used in each separate
model fit are listed in Table~\ref{diff_ind_tab}.  We substitute each
alternative index separately to asses the effect due to that
individual index.  We also examine some combinations of indices,
specifically a ``balmer'' model, in which we fit using the average of
H$\beta$, H$\gamma_F$, and H$\delta_F$, an ``all'' model, in which all
lines are included in the fit, and a ``high-z'' model, which simulates
the effect of applying EZ\_Ages to higher redshift data, where only
the bluer indices ($\lambda < 4400$ \AA) may be available because the
indices at longer wavelengths are redshift out of the optical
spectrum.  Results from the fitting process are provided in
Table~\ref{diff_fit_tab} and displayed in
Figures~\ref{vary_fit_NGC6441} and \ref{vary_fit_M67}, where residuals
relative to the age and metal abundances obtained with the standard
index set are plotted for the various index combinations.

Looking first at NGC~6441, Figure~\ref{vary_fit_NGC6441} shows that,
for the vast majority of the index combinations, ages and metal
abundances vary by less than $\pm0.1$ dex, indicating that EZ\_Ages
and the S07 models have attained an outstanding degree of consistency.
There are a few exceptions, though, which occur for the following
index combinations: 1) When G4300 is used as a carbon abundance
indicator in place of C$_2$4668 (panel c, triangles); 2) When higher
order Balmer lines are used as age indicators (panel a); 3) When
higher order Balmer lines are used as age indicators and Fe4383 is
used as the iron-abundance indicator as in the ``high-z'' model (panel
d, triangles) or when the higher order Balmer lines and Fe4383 are
averaged in with the other lines as in the ``all'' model (panel d,
squares).  In case 1, the G4300-based [C/Fe] and [N/Fe] values differ
from those obtained from the standard indices by --0.14 and +0.18,
respectively.  In case 2, ages tend to be younger when higher-oder
Balmer lines are used in place of H$\beta$, by up to $\sim$ --0.2 dex.
Finally, case 3 is similar to case 2, except that [Fe/H] is too high,
and consequently all the other abundances are too low.

The difference found in case 1 when replacing C$_2$4668 by the G4300
index is due to the different sensitivities of the two indices to the
ratio between the abundances of carbon and oxygen (C/O).  The two
indices are sensitive to variations of the oxygen abundance because of
details of the molecular dissociation equilibrium in the atmospheres
of cool stars.  Of all molecules in whose formation carbon and oxygen
take part, carbon monoxide (CO) is the hardest one to break, because
it has the highest dissociation energy.  Therefore, at the
temperatures prevalent in the atmospheres of G and K stars, most
available free carbon and oxygen atoms are locked in CO.  As a result,
variations in the abundance of oxygen, usually the most abundant of
the two species, have a strong influence on the amount of carbon that
is free to form other molecules, such as C$_2$ and CH, which are
responsible for the vibrational bands measured by the C$_2$4668 and
G4300 indices, respectively.  Therefore, an increase in the abundance
of oxygen tends to provoke a decrease in the strength of those
molecular bands.

While the concentration of the C$_2$ molecule depends quadratically on
the abundance of carbon, that of CH depends only linearly on that
abundance.  As a result, C$_2$4668 is far more sensitive to carbon
than to oxygen, while the the sensitivity of the G4300 index to those
two abundances is approximately the same (though for both indices, the
sensitivity to carbon and oxygen abundances have opposite signs).
Although this makes the G4300 index a more uncertain indicator of
carbon abundance, it raises the possibility that a combination of the
two indices may be used to constrain the abundance of oxygen.  We
verified that possibility by raising the input oxygen abundance in the
models from [O/Fe]~=~0 to +0.3.  As a result, the carbon and nitrogen
abundances inferred from use of C$_2$4668 and G4300 indices agreed to
within 0.05 dex.  Furthermore, the ages and abundances inferred for
the O-enhanced models were not substantially different from the
results using the [O/Fe]~=~0 model\footnote[3]{The changes in age and
abundance determinations from increasing [O/Fe] to $+0.3$ dex are
small because only the effect on the atmospheric line strengths are
taken into account.  The effect of enhancing [O/Fe] in the isochrone
are not included because reliable O-enhanced isochrones have not yet
been incorporated into the S07 model.  An O-enhanced isochrone would
result in younger age measurements without significantly changing the
inferred abundances.  S07 shows that model grids with
$\alpha$-enhanced isochrones move along lines of constant [Fe/H] in
index-index plots.  The result is that the inferred ages are different
when using $\alpha$-enhanced isochrones, but that the abundance
determinations are relatively unchanged.}: [N/Fe] increased by $+0.12$
dex, while the ages inferred from H$\gamma_F$ and H$\delta_F$
increased by 0.6 Gyr.  The values of H$\beta$ age, [Fe/H], [Mg/Fe],
and [Ca/Fe] showed negligible change ($< 0.1$ Gyr or $< 0.05$ dex).
Interestingly, the oxygen abundance in NGC~6441 stars ranges from
[O/Fe] = --0.05 to 0.34 \citep{gra06}, with stars in the upper end of
the interval having ``normal'' oxygen abundances, and likely being
more numerous and dominating the cluster light.  Therefore, the value
we obtained from the cluster's integrated light is in good agreement
with the known cluster abundance, which makes it possible that oxygen
abundances might be inferred from a combination of the C$_2$4668 and
G4300 indices.  This will be further investigated in a forthcoming
paper.

The dependence in case 2 of the resulting age on the Balmer line
adopted is not surprising, given that NGC~6441 is characterized by the
presence of blue HB stars (\citealt{ric97}, \citealt{bus07}).  Because
these stars have early-F and A spectral types, they tend to strengthen
Balmer lines in the integrated spectra.  That effect is obviously more
important at bluer wavelengths, where the contribution by early-F and
A stars to the integrated light is the greatest.  Therefore,
higher-order Balmer lines tend to be more strongly affected.  As a
result, when models that do not take into account blue HB stars are
compared with the data, they tend to infer systematically younger
ages, and more so according to higher order Balmer lines (see
discussion in S07).  \citet{sch04b} have in fact shown that this
effect can be used to constrain the morphology of the HB of globular
clusters, solely on the basis of Balmer line strengths in their
integrated spectra.  In particular, the ratio between H$\delta_F$ and
H$\beta$ was shown to be very sensitive to the presence of blue HB
stars.

Before discussing case 3, where Fe4383 and H$\delta_F$ are used in the
fitting process, let us look at the results of the consistency test
for M~67.  Here again there is agreement to within $\pm0.1$ dex for
any combination of indices used, with the exception of combining
Fe4383 and H$\delta_F$.  The three Balmer lines all give consistent
results for M~67 (panel a), because M~67 does not contain blue HB
stars and because the integrated spectrum of M~67 used in the this
analysis has been carefully constructed to exclude blue stragglers
\citep{sch04a}.  Unfortunately, there is a problem (as yet not
well-understood) with the G4300 measurements for this cluster
\citep{sch04a}, thus we cannot use M~67 to test our hypothesis about
the effect of oxygen abundance on G4300 and C$_2$4668.

Indeed, the only inconsistency in the abundance determinations for
M~67 is in the ``high-z'' model, in which Fe4383 and H$\delta_F$ are
used in the fitting process (case 3 from above).  Here, as with
NGC~6441, we not only derive younger ages, but the resulting iron
abundance is higher by $\sim +0.24$ dex.  Consequently, all the other
elemental abundances are found to be somewhat low.  For M~67, we
cannot invoke blue HB stars to explain this discrepancy.
Interestingly, substituting H$\delta_F$ or Fe4383 into the standard
index set {\it individually} (triangles in panel a and squares in
panel b) does not result in large discrepancies from the standard
model.  How then can we understand the dramatic differences seen when
both H$\delta_F$ and Fe4383 are used in the modelling process?

\begin{figure}
\epsscale{0.9}
\plotone{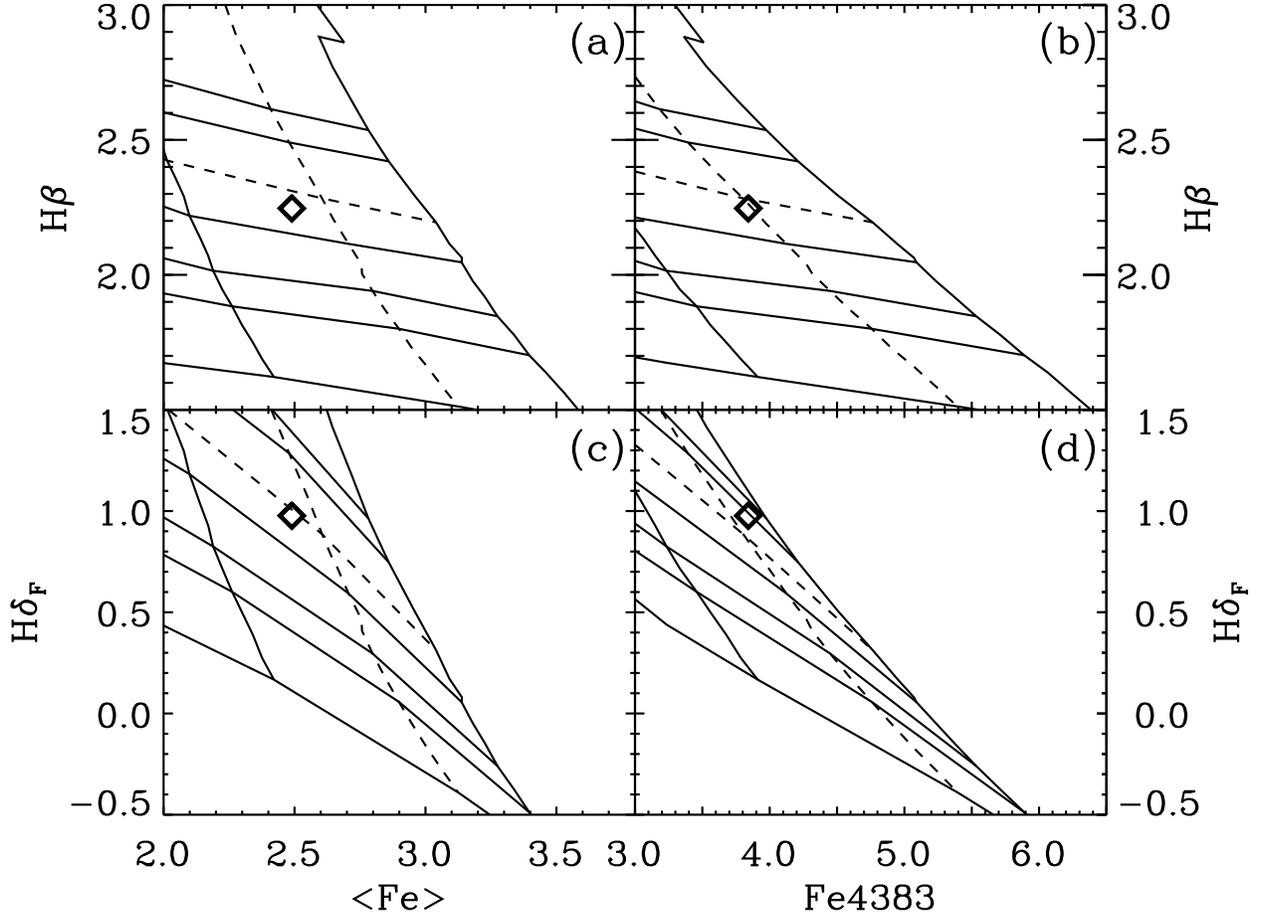}
\caption{Solar abundance model grids showing the difficulty of
  measuring age and [Fe/H] from Fe4383 and H$\delta_F$.  Lines of
  constant age run nearly horizontally and show (from top to bottom)
  2.5, 2.8, 3.5, 5.0, 7.0, 8.9, and 14.1 Gyr models.  Lines of
  constant [Fe/H] run nearly vertically and show (from left to right)
  $-0.4$, 0.0, and $+0.2$ dex models.  The 3.5 Gyr and solar abundance
  model are shown with dashed lines for reference.  Measured index
  values for M~67 are overplotted as the diamond.  The standard
  indices (panel a) produce model grids with lines of constant age and
  constant [Fe/H] nearly perpendicular.  Line of constant [Fe/H] are
  more sloped from the vertical when using Fe4383 instead of
  $\langle$Fe$\rangle$ (panel b).  Similarly, lines of constant age
  are more sloped from the horizontal when using H$\delta_F$ instead
  of H$\beta$ (panel c).  Substituting one or the other of these
  indices for the standard set causes only small offsets from the
  standard results (see Figure \ref{vary_fit_M67}).  However, the
  combination of Fe4383 and H$\delta_F$ (panel d) results in model
  grids that collapse down on one another, making the results much
  more sensitive to small errors or zeropoint uncertainties in either
  the models or the data.  Because of the near-degeneracy of the
  models in Fe4383-H$\delta_F$ space, the combination of these two
  indices can give results substantially different from the age and
  [Fe/H] measured using the standard index set.
  }\label{red_blue_grids}
\end{figure}

Figure \ref{red_blue_grids} shows solar abundance model grids from the
S07 model.  The M~67 data are overplotted as diamonds.  In panel a, we
see that H$\beta$ and $\langle$Fe$\rangle$ produce model grids where
lines of constant age and constant [Fe/H] are nearly perpendicular.
This is one of the reasons this index combination is so useful for age
and [Fe/H] determinations.  When Fe4383 is substituted for
$\langle$Fe$\rangle$ (panel b), the lines of constant [Fe/H] are less
nearly vertical, lessening the diagnostic power of the index-index
grid.  Likewise, when H$\delta_F$ is substituted for H$\beta$ (panel
c), the lines of constant age are less nearly horizontal, again
constraining the age and [Fe/H] determinations.  This is because, in
the blue region, line crowding is higher, making H$\delta_F$ more
sensitive to metallicity.  Moreover, the blue spectral region is more
strongly affected by warm stars than the red, making Fe4383 more
sensitive to age than its redder counterparts.  Notice, however, that
there is still enough spread in the grids in panels b and c that small
errors or zero-point uncertainties in the data or the models do not
result in substatially different age and [Fe/H] measurements for M~67.

Now consider panel d.  When H$\delta_F$ and Fe4383 are combined in the
fitting process, the model grid lines are very far from perpendicular
and the model space collapses down toward the familiar age-metallicity
degeneracy.  Here, small errors or zero-point uncertainties result in
substantially different age and [Fe/H] measurements.  Even the slight
metallicity difference seen between panels a and b becomes a
substantial discrepancy in panel d.  This raises warning flags for
using these indices to determine age and [Fe/H].  In particular, we
note that Fe4383 appears to {\it always} give slightly higher values
of [Fe/H] in the data presented both here and in ongoing work by the
authors, suggesting that there is some inaccuracy in the modelling of
this index.  The fact that the discrepancy appears in the case of
M~67, whose abundances are known to be solar-scale, suggests that this
may be a zeropoint problem in the models, rather than a problem in
calculating the abundance sensitivities of this index.  Combined with
the increased degeneracy of the H$\delta_F$-Fe4383 index-index space,
this means that EZ\_Ages should be used cautiously when determining
stellar population parameters in higher redshift objects where
H$\beta$ is not available, and that the need for high $S/N$ data is
particularly acute in this regime.  Even when the problem with Fe4383
is solved, the degeneracy of H$\delta_F$-Fe4383 index-index space
means that small errors or zeropoint uncertainties in data can result
in errors in the determination of age and [Fe/H] from these indices.
This likely applies to all stellar population models, not just those
presented in this work and S07.

In summary, we conclude that EZ\_Ages is characterized by a remarkable
degree of consistency, in that ages and metal abundances inferred from
different index sets agree in the vast majority of cases to within
$\pm0.1$ dex.  Two of the inconsistencies we found, namely, that
different Balmer lines indicate different ages, and different
carbon-abundance indicators indicate different carbon abundances, can
in fact be potentially explored to extract even more information from
the integrated spectra of stellar populations.  While blue Balmer
lines were shown to be useful to indicate the presence of blue HB
stars in globular clusters \citep{sch04b}, the carbon index
discrepancy can potentially be used to constrain the abundance of
oxygen.  The latter effect will be further investigated in a
forthcoming paper.  The third inconsistency is due to the fact that
the combined H$\delta_F$-Fe4383 index-index space is substantially
more degenerate than H$\beta$-$\langle$Fe$\rangle$ space, making
accurate age and [Fe/H] determinations from these indices difficult.
This is a matter of concern in integrated studies of globular clusters
\citep{sch04b} and higher redshift galaxies \citep{sch06}, for which
only these bluer lines may be available.

\section{Comparison with Abundance Modeling of
  Galaxies}\label{compare_galaxies} 

Having performed the reality and consistency checks above, we now turn
to comparisons between EZ\_Ages results and those coming from
application of other models in the literature.  We focus on the models
by \citet[hereafter TMB03]{tho03}, which also make predictions for
variable abundance patterns.  The TMB03 models also attempt to match
non-solar abundance patterns by fitting Lick indices in unresolved
stellar population spectra.  These include both solar-scaled and
$\alpha$-enhanced models, in which the set of $\alpha$ elements N, O,
Mg, Ca, Na, Ne, S, Si, and Ti are slightly enhanced, while the iron
peak elements Cr, Mn, Fe, Co, Ni, Cu, and Zn are significantly
depressed, in order to vary [$\alpha$/Fe] at fixed total metallicity
[Z/H].  In these models, [C/Fe] is fixed at solar.

\citet[hereafter T05]{tho05} use the TMB03 models to estimate ages,
[Z/H], and [$\alpha$/Fe] for a sample of 124 early-type galaxies,
based on measurement of the Lick indices H$\beta$, $\langle$Fe$\rangle$, and
Mg~$b$.  To compare the stellar population modeling results of using
EZ\_Ages and the S07 models to those using TMB03, we have run EZ\_Ages
on the sample of 124 galaxies presented in T05, using the values of
H$\beta$, $\langle$Fe$\rangle$, and Mg~$b$ given in their Table 2.  With these
indices, EZ\_Ages can determine the SSP age, [Fe/H], and [Mg/Fe]
values of the sample galaxies and compare these results with those of
the TMB03 models used in T05.

Although the details of the model fitting procedure differ, T05
perform a very similar analysis to that of EZ\_Ages, determining age,
[Z/H], and [$\alpha$/Fe] from the same Lick indices that are the
standard set EZ\_Ages uses to calculate age, [Fe/H], and [Mg/Fe].  The
models used in T05 use the same \citet{kor05} index sensitivity
functions as S07 and should therefore show the same index strength
variations when the abundances are modified.  One practical difference
is that the TMB03 models are cast in terms of [Z/H], rather than
[Fe/H].  In order to compare results from the two models, we convert
T05's estimated [Z/H] into [Fe/H] using the conversion proposed by
TMB03: [Fe/H] = [Z/H] $- 0.94$ [$\alpha$/Fe] (TMB03).

The TMB03 models assume that all $\alpha$-elements track Mg.  EZ\_Ages
follows a similar convention, by default setting Na, Si, and Ti to
track Mg in the modeling process, while Cr follows Fe.  A significant
advantage of EZ\_Ages and the S07 models is that they also track the
[C/Fe], [N/Fe], and [Ca/Fe] abundance ratios using C, N, and
Ca-sensitive Lick indices, rather than making assumptions about how
these elements vary.  However, these differences should not affect the
comparison of age, [Fe/H], and [$\alpha$/Fe] results presented here,
as H$\beta$, $\langle$Fe$\rangle$, and Mg~$b$ are predominantly
sensitive to age, Fe and Mg only (see Table \ref{indextable}), and are
thus treated the same by both models through the \citet{kor05}
sensitivity functions.

\begin{figure}
\epsscale{1.0}
\plotone{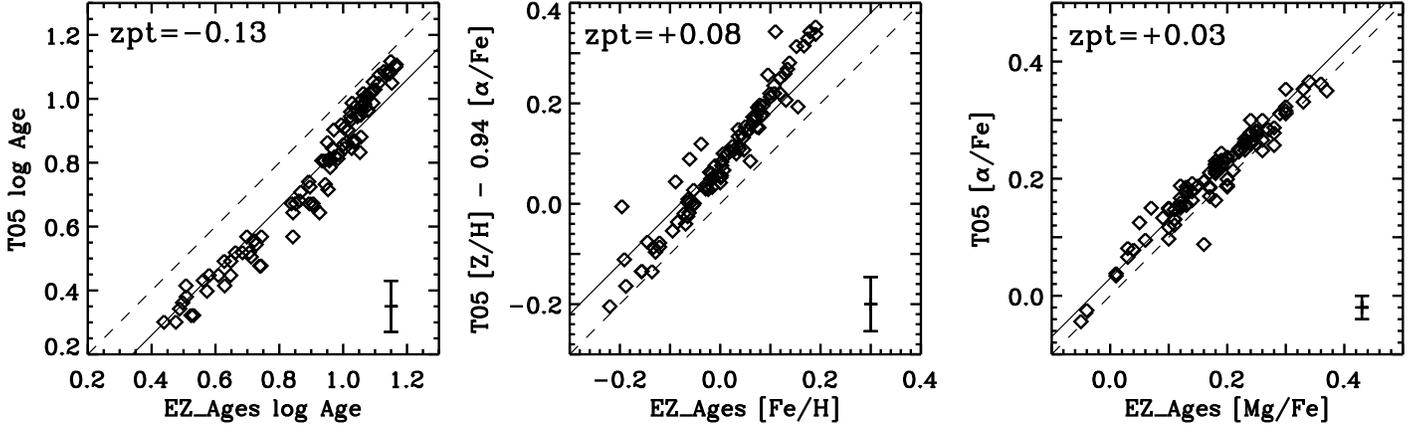}
\caption{Comparison of $\log$ Age, [Fe/H], and [$\alpha$/Fe] results
  from T05 and from EZ\_Ages.  The galaxy sample is that of T05, and
  the stellar population parameters are from SSP fits to H$\beta$,
  $\langle$Fe$\rangle$, and Mg~$b$ for each set of models.  T05 models
  are cast in terms of [Z/H], which we compare to EZ\_Ages values of
  [Fe/H] using the conversion given in TMB03 for their
  $\alpha$-enhanced mixture: [Fe/H] = [Z/H] - 0.94 [$\alpha$/Fe].  We
  compare their values of [$\alpha$/Fe] to the EZ\_Ages results for
  [Mg/Fe], since T05 use the Mg~$b$ line as their $\alpha$-enhancement
  indicator.  Dashed lines show the one-to-one relation.  Solid lines
  show the best fit one-to-one slope allowing for a zeropoint offset.
  The size and direction of the zeropoint offset are indicated in the
  top left of each panel.  Error bars in the lower right corner of
  each panel indicate the median observational errors from T05.
  Results for [$\alpha$/Fe] are very consistent between the two
  models.  T05 find ages which are younger by $\sim 35$\% and [Fe/H]
  values which are higher by $\sim 0.08$ dex than the EZ\_Ages
  results.  Also, the slope of the [Fe/H] comparison differs somewhat
  from the one-to-one relation.  These zeropoint differences, as well
  as the slope difference in [Fe/H], can be explained by differences
  in the model ages, which affect the [Fe/H] estimates, but have
  relatively little effect on [$\alpha$/Fe].  The age zeropoint
  difference between the models is somewhat larger than the zeropoint
  uncertainty in the S07 models (the age zeropoint of the models used
  in T05 is uncalibrated), suggesting that the T05 analysis may
  slightly underestimate the age of the sample galaxies.  See text for
  details.  }\label{compare_thomas}
\end{figure}

The results of fitting the T05 galaxies using EZ\_Ages are shown in
Figure \ref{compare_thomas}.  In each panel, the dashed line shows a
one-to-one relation and the solid line shows the best fitting line to
the data while keeping the one-to-one slope fixed.  The resulting
zeropoint offsets between the model fits are indicated in the upper
left corner of each panel.  In general, there is good agreement
between the results coming from application of the two models to the
same data set.  The scatter is small and the zeropoint offsets are
modest.  The values of [$\alpha$/Fe] agree to within a very small
offset.  The ages estimated in T05 are typically $\sim 25$\% (0.13 dex)
younger than those we find with EZ\_Ages, while the T05 values of
[Fe/H] are $\sim 0.08$ dex higher.

In addition to the zeropoint offsets in the age and [Fe/H]
comparisons, the two different analyses do not truly follow a
one-to-one slope for either parameter: the [Fe/H] relation is
significantly steeper than the unity relation, while the age
comparison shows some curvature.  At old ages ($\log$ age $\simgreat
1.0$ as estimated by EZ\_Ages) the slope is very nearly one-to-one,
and the agreement between the two age estimates is much better, with a
zeropoint offset of 15\% ($-0.07$ dex) between the T05 ages and those
from EZ\_Ages.  At intermediate ages ($0.8 < \log \mbox{age} < 1.0$ as
estimated by EZ\_Ages) the relation is steeper than unity, while at
young ages ($\log \mbox{age} \simless 0.8$) the relation flattens out
again, although at a substantial zeropoint offset.  At intermediate
and young ages, the T05 ages are 33\% (0.17 dex in log age) younger.
The origin of the curvature in the age relation is unclear.  In any
case, age agreement for old galaxies is very good (within 15\%) and is
still within 33\% for all galaxies down to the youngest SSP ages
estimated in the T05 sample.

Based on the cluster comparison of \S\ref{real} and Figure
\ref{comp_lit}, the zeropoint uncertainties in the S07 models are
roughly 0.07 dex in $\log$ age, 0.1 dex in [Fe/H], and 0.05 dex in
[Mg/Fe].  The differences between the T05 and the EZ\_Ages results are
therefore within the zeropoint uncertainties of the models, with the
exception of the age estimates for galaxies younger than 10 Gyr.  For
these galaxies, the T05 results are younger by more than twice the
indicated zeropoint uncertainties in the S07 and EZ\_Ages models.  The
best way to resolve this discrepancy is to compare the models with
data for metal-rich, intermediate-age clusters.  While the models by
S07 have been shown in the previous section and in Schiavon (2007) to
match the data for M~67 ( $\sim$ 3.5 Gyr-old and solar metallicity),
more data are clearly needed to better constrain the models in this
crucial age/metallicity regime.

The non-unity slope of the [Fe/H] comparison also deserves further
investigation.  While the differences are negligibly small in the
solar metallicity regime, they climb up to 0.2 dex at the
high-metallicity end.  For the most metal-rich galaxies, EZ\_Ages
obtains [Fe/H] $\sim$ 0.2, and T05 find galaxies to be more metal-rich
by $\sim$ 0.2 dex.  As in the case of age determinations, we again
find ourselves lacking data that could help decide between the two
model sets, given the non-existence of integrated spectroscopy for
super-metal-rich clusters.  It would be interesting, however, to
compare results from the two models by confronting them with cluster
data in a regime for which such data are available, such as moderately
metal-poor clusters.

\begin{figure}
\epsscale{1.0}
\plotone{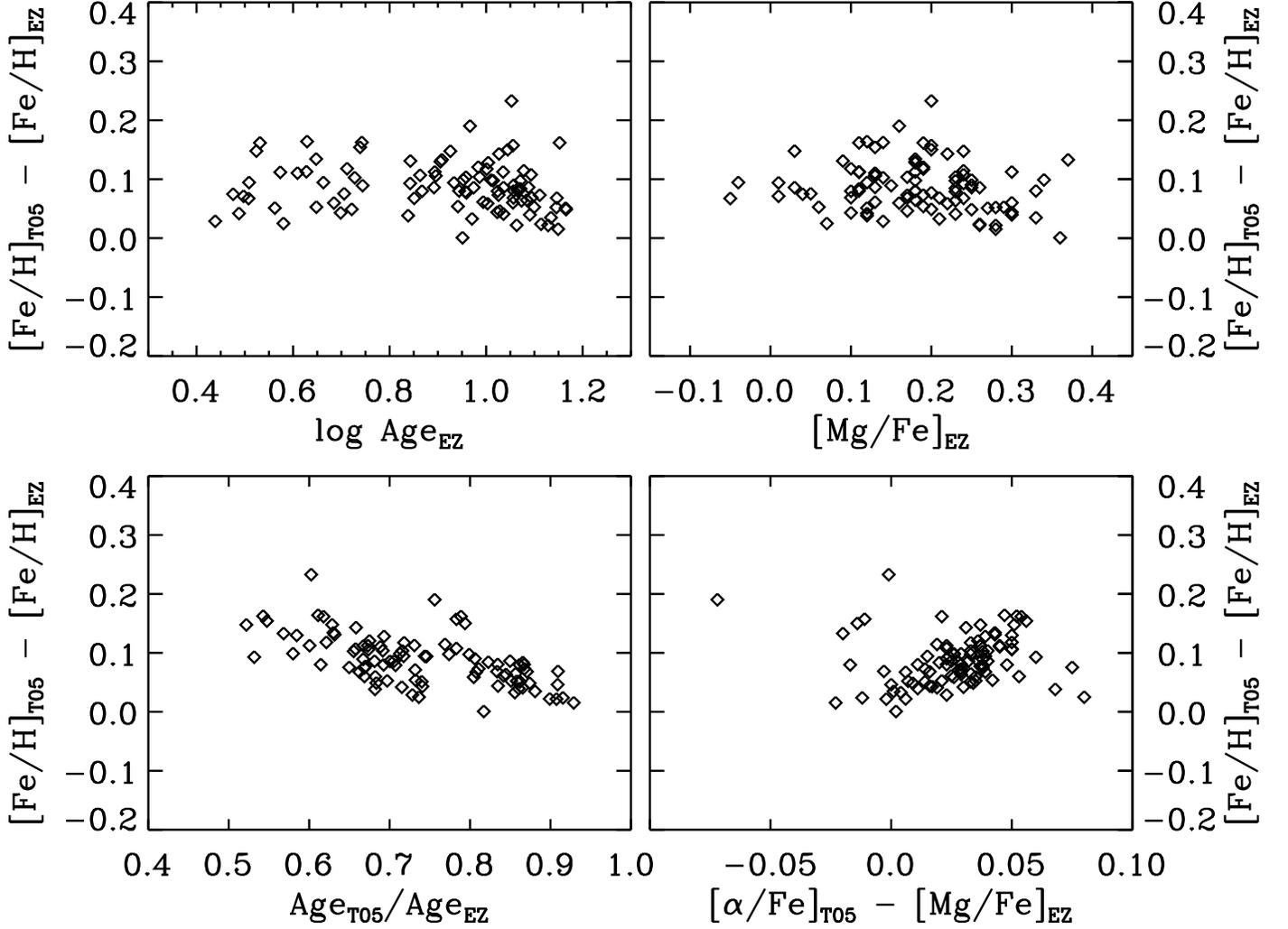}
\caption{{\it Top:} Difference between [Fe/H] estimates for the T05
  galaxy sample from EZ\_Ages and from T05, as a function of stellar
  population age and [Mg/Fe].  {\it Bottom:} Difference in [Fe/H] as a
  function of differences in stellar population age and [$\alpha$/Fe].
  The differences between T05 and EZ\_Ages estimates of [Fe/H] are
  most strongly correlated with differences in the age results between
  the two models (lower left panel).  This suggests that the non-unity
  slope of the [Fe/H] comparison relation in Figure
  \ref{compare_thomas} is due to differences in the age estimates and
  arises from correlated errors (see discussion in \S\ref{errors}).
  The non-orthogonality of model grids in the index-index diagrams (as
  in Figure \ref{grids}) results in correlated errors such that
  underestimating the age of a population will result in
  overestimating the corresponding value of [Fe/H].
  }\label{thomas_slope}
\end{figure}

In Figure \ref{thomas_slope}, we plot the difference between the
values of [Fe/H] estimated in T05 and by EZ\_Ages as a function of
various other properties of the stellar populations .  The top panel
shows the [Fe/H] differences as a function of SSP age and
$\alpha$-enhancement.  The differences in the [Fe/H] estimates do not
appear to be strongly correlated with either SSP age or
$\alpha$-enhancement.  In the lower panel, we show the differences in
[Fe/H] as a function of the differences in the estimated ages and
$\alpha$-enhancements.  The [Fe/H] differences are correlated with
differences in the age estimates between the two different models.
Furthermore, the correlation is in the direction expected from the
correlated errors in the index-index plots: where T05 obtain
substantially younger ages than EZ\_Ages, they also find higher
[Fe/H].  The difference in the age estimates by the different models
therefore naturally explains the observed differences in [Fe/H], both
the zeropoint shift and the non-unity slope of the [Fe/H] comparison.
Estimates of [$\alpha$/Fe] are substantially less biased by a
zeropoint error in the age.  That is because the Mg~$b$ and
$\langle$Fe$\rangle$ indices have very similar age-dependence, so that
the zero-point uncertainties in the parameters determined by these two
indices ([Mg/H] and [Fe/H]) cancel out.

From this analysis, it is clear that the TMB03 models and EZ\_Ages
give fairly consistent results, with a relatively small disagreement
in the age zeropoints of the models which produces modest
disagreements in [Fe/H].  Abundance ratios such as [$\alpha$/Fe] are
much less sensitive to zeropoint offsets.  The comparison of EZ\_Ages
results with ages from CMD fitting of Galactic clusters
(\S\ref{compare_clusters}) indicates that the age zeropoint in
EZ\_Ages is correct to within 0.07 dex in $\log$ age.  The [Z/H] and
[$\alpha$/Fe] results from TMB03 models have been compared with
cluster data \citep{mar03,tho03}, but age estimates have not been
independently tested, therefore it is difficult to say what the
zeropoint uncertainties in the age determinations for these models
should be.  We therefore conclude that EZ\_Ages and the S07 models
give results in good agreement with the TMB03 models (at least in the
high-metallicity regime tested in this analysis of galaxy spectra),
but that at intermediate ages (age $< 10$ Gyr) the TMB03 SSP ages may
underestimate the true SSP ages by as much as 33\% (0.17 dex in $\log$
age).

A more general result from this comparison is that the sequential grid
inversion algorithm presented in this paper, combined with the SSP
models of S07, does an excellent job of reproducing [Mg/Fe] abundances
obtained by other groups with other models.  Although the EZ\_Ages
abundance results for [C/Fe], [N/Fe], and [Ca/Fe] in unresolved
stellar populations cannot be compared to other models (since no other
models exist for the comparison), the modeling process for fitting
these other abundances is the same as that used to fit [Mg/Fe] and the
success of this process for [Mg/Fe] bodes well for other abundances.
The test presented in this section, combined with the stringent tests
with cluster data described in \S\ref{compare_clusters}, are
reassuring evidence that this modeling process is reasonable and can
be used to provide a {\it quantitative} assessment of multiple
elemental abundances.

\section{Conclusions}\label{conc}

In this paper, we have presented a methodology for measuring SSP ages,
[Fe/H], and individual abundance ratios [Mg/Fe], [C/Fe], [N/Fe], and
[Ca/Fe] for unresolved stellar populations.  We do this by exploiting
the different sensitivities of various Lick indices to a variety of
elemental abundances, and the ability of the S07 stellar population
models to accurately model the Lick indices for a wide range of input
abundance patterns.  The algorithm presented here has been implemented
in the IDL code package EZ\_Ages, which is available for download and
general use.

We have subjected the modeling process described here to numerous
rigorous tests and comparisons, with the following results:
\begin{list}{}{}
\item[1.] {\it Comparison with Galactic cluster data:}
\begin{list}{}{}
\item[a.] {\it Ages:} EZ\_Ages age estimates reproduce the results of
  cluster CMD fitting to within 0.15 dex for all clusters, with the
  exception of NGC~6121, whose Balmer line strengths are affected by a
  blue HB.  For two out of three of the remaining clusters, the
  EZ\_Ages age is within 0.1 dex of the CMD result.
\item[b.] {\it [Fe/H] and [Mg/Fe]:} EZ\_Ages estimates match the
  results from high-resolution spectroscopy of individual cluster
  members to within 0.1 dex for some of the clusters, and to within
  0.2 dex for all clusters (with the exception of the disagreement
  between our value of [Fe/H] for NGC~6528 and the results of
  \citealt{car01}---we are in $\sim \pm0.1$ dex agreement with the
  results of \citealt{zoc04} and \citealt{bar04} for [Fe/H] for this
  cluster).
\item[c.] {\it [C/Fe] and [N/Fe]:} For the only two clusters where C
  and N abundances are available (47~Tuc and M~67), EZ\_Ages results
  match those from high-resolution spectroscopy of individual cluster
  members to within 0.1 dex.  For NGC~6121, we find suspiciously low
  values, that might indicate a problem at the low metallicity end of
  our models.  Therefore, we strongly caution users against using
  EZ\_Ages to determine these abundances for systems with [Fe/H]
  $\simless -1.0$.
\item[d.] {\it [Ca/Fe]:} EZ\_Ages results are within $\sim 0.1$ dex
  for all clusters except NGC~6121.  The latter is certainly related
  to the problem mentioned above regarding [C/Fe] and [N/Fe] in the
  metal-poor regime, so we do not recommend use of EZ\_Ages for
  [Ca/Fe] determinations for systems with [Fe/H] $\simless -1.0$.
\end{list}
\item[2.] {\it Consistency test performed on the cluster NGC~6441:}
  using different combinations of Lick indices in the fitting process
  yields results that are consistent to within $\pm 0.1$ dex for all
  stellar population parameters except for the following cases:
\begin{list}{}{}
\item[a.] Using G4300 instead of C$_2$4668 to fit [C/Fe] results in
  values of [C/Fe] and [N/Fe] that differ by $-0.14$ and $+0.18$ dex,
  respectively.  This difference is probably due to differences in the
  index responses to O, which are not explored in this work.
\item[b.] Using H$\gamma_F$ or H$\delta_F$ instead of H$\beta$ as an
  age indicator results in younger ages for the cluster NGC~6441 due
  to the effect of the cluster's blue HB.  This discrepancy does not
  exist for M~67, which does not have a blue HB, and whose integrated
  spectrum is constructed to exclude blue HB stars.
\item[c.] Using the bluest Balmer and Fe lines available, H$\delta_F$
  and Fe4383, can result in age and [Fe/H] measurements that are
  substantially different than those determined using the standard set
  of indices, due to the increased degeneracy of the model space for
  these indices.  This poses a problem for high redshift studies of
  stellar populations, where often only the bluest lines are available
  for analysis.
\end{list}
\item[3.] {\it Error analysis test:} EZ\_Ages implements an algorithm
  that attempts to propagate errors through the modeling process in
  an efficient way that takes into account the dependence of abundance
  fitting on other parts of the population fitting process.  Monte
  Carlo simulations show that this simplified error estimation does an
  excellent job of matching the true errors in the population
  parameters due to measurement errors in the Lick indices.
\item[4.] {\it Comparison with \citet{tho05} results:}
\begin{list}{}{}
\item[a.] Results for [$\alpha$/Fe] are consistent between EZ\_Ages
  results and those using TMB03 models.  
\item[b.] Age and [Fe/H] comparisons show little scatter.  There are
  zeropoint offsets of 15\%--33\% (0.07--0.17 dex) in age and 0.08 dex
  in [Fe/H] between the two analyses, and a non-unity slope in the
  [Fe/H] comparison.  The age differences are small for the oldest
  (age $> 10$ Gyr) galaxies and increase for younger galaxies.  The
  [Fe/H] effects seem to be a natural result of the age differences.
  The S07 and EZ\_Ages age zeropoints are shown here to be good to
  within 0.07 dex, while the TMB03 age zeropoints are untested.  This
  suggests that the ages derived in T05 may be too young by up to 33\%
  (0.17 dex) for intermediate age galaxies (age $< 10$ Gyr).
\item[c.] Absolute estimates of abundance ratios are more reliable
  than absolute estimates of age or [Fe/H] because zeropoint
  uncertainties tend to cancel out.
\item[d.] The sequential grid inversion algorithm method is proven to
  work for estimating [Mg/Fe].  Since this method is the same as the
  one used to determine other abundance ratios, those estimates are
  likely to also be reliable, as demonstrated in the comparison with
  cluster data.
\end{list}
\end{list}

Overall, EZ\_Ages and the S07 SSP models do an excellent job of
fitting the SSP age, [Fe/H], and abundance ratios [Mg/Fe], [C/Fe],
[N/Fe], and [Ca/Fe] for unresolved stellar populations.  The small
zeropoint uncertainties in age and [Fe/H] estimates illustrated by
comparisons with Galactic cluster data demonstrate that absolute
estimates can be made for these quantities with high $S/N$ data.
Absolute estimates of elemental abundances are robust to the (small)
zeropoint uncertainties.  EZ\_Ages and the S07 models therefore make
it possible, for the first time, to perform a quantitative assessment
of multiple individual elemental abundances from medium-resolution
spectra of unresolved stellar populations.  Furthermore, the abundance
fitting process can be run in an automated way on large data sets.
With these innovations, stellar population analysis is better
positioned than ever before to address the task of unravelling the
star formation histories of stellar systems.

\acknowledgements{The authors wish to thank an anonymous referee for
  excellent and thought-provoking comments that improved the quality
  of this work.  They wish to thank Andreas Korn for valuable
  information concerning use of the Korn et al. (2005) sensitivity
  tables.  This work is supposrted by NSF grants AST 00-71198 and AST
  05-07483.  R. P. S. is supported by Gemini Observatory, which is
  operated by the Association of Universities for Research in
  Astronomy, Inc., on behalf of the international Gemini partnership
  of Argentina, Australia, Brazil, Canada, Chile, the United Kingdom,
  and the United States of America.}

\begin{deluxetable}{lcc}
\tabletypesize{\scriptsize}
\tablecaption{Lick Indices in the S07 SSP Models. \label{indextable}}
\tablewidth{0pt}
\tablehead{
\colhead{Index} &
\colhead{KMT Sensitivity\tablenotemark{*}} &
\colhead{SWB Sensitivity\tablenotemark{*}}
}

\startdata
\multicolumn{3}{c}{Balmer Indices} \\[3pt]
\tableline \\
H$\delta_A$     &Fe, C$^-$  &Fe, C, V          \\
H$\delta_F$     &Fe, Mg$^+$ &Fe, C, Si         \\
H$\gamma_A$     &C, Fe, Mg  &C, Ti, (Mg)       \\
H$\gamma_F$     &C, Fe      &C, (Si), (Mg)    \\
H$\beta$        &...        &(Ni)             \\
\tableline \\
\multicolumn{3}{c}{Fe Indices} \\[3pt]
\tableline \\
Fe4383          &Fe, Mg, C$^-$ &C, Fe        \\
Fe5015          &Ti, Mg, Fe    &C, Fe, (Ti)  \\
Fe5270          &Fe            &C, (Fe)      \\
Fe5335          &Fe            &Fe, C        \\
\tableline \\
\multicolumn{3}{c}{Mg Indices} \\[3pt]
\tableline \\
Mg$_2$          &Mg, C     &Mg, C, Fe    \\
Mg~{\it b}      &Mg, Fe, C &Mg, Fe, (Cr) \\
\tableline \\
\multicolumn{3}{c}{C, CH, and CN Indices} \\[3pt]
\tableline \\
CN$_1$             &C, N, O      &C, N, O      \\
CN$_2$             &C, N, O      &C, N, O      \\
G4300              &C, O, Fe$^+$ &C, Fe, Ti    \\
C$_2$4668          &C, O         &C, O, (Si)   \\
\tableline \\
\multicolumn{3}{c}{Ca Indices} \\[3pt]
\tableline \\
Ca4227          &Ca, C  &Ca, O, (CN)  
\enddata
\tablecomments{Index sensitivities as given by \citet[KMT]{kor05} and
  \citet[SWB]{ser05}.  Only the top three element sensitivities are
  given, and only those with significance above 1$\sigma$ in the model
  spectra.  For the KMT sensitivities, only those of turn-off
  and giant branch stars are shown, as these two components comprise
  90\% of the light in the S07 models.  For the SWB sensivities, those
  under 2$\sigma$ are shown in parentheses.}
\tablenotetext{*}{In addition to the reported sensitivities for
  individual element abundances, all indices listed here vary with
  total metallicity.}
\tablenotetext{+}{Indicates sensitivities that only appear at high
  metallicity.}
\tablenotetext{-}{Indicates sensitivities that only appear at low
  metallicity.}

\end{deluxetable}

\begin{deluxetable}{lccccc|cc}
\tabletypesize{\scriptsize}
\tablecaption{Selected Lick Indices for Clusters\label{cluster_indices}}
\tablewidth{0pt}
\tablehead{
\colhead{} &
\colhead{NGC~6121} &
\colhead{47~Tuc} &
\colhead{NGC~6441} &
\colhead{NGC~6528} &
\colhead{M~67} &
\colhead{GC Err$^{*}$} &
\colhead{M~67 Err} 
}
\startdata
H$\delta_F$    & 2.338 & 0.682 & 0.926 & 0.298 & 0.977  &0.117 &0.090\\
H$\gamma_F$    & 1.398 &-0.660 &-0.148 &-1.308 &-0.324  &0.215 &0.050\\
H$\beta$       & 2.280 & 1.602 & 1.804 & 1.572 & 2.246  &0.083 &0.090\\
Fe4383	       & 1.267 & 2.491 & 3.048 & 4.666 & 3.841  &0.267 &0.200\\
Fe5270	       & 1.211 & 1.913 & 2.128 & 2.799 & 2.647  &0.142 &0.090\\
Fe5335	       & 0.984 & 1.685 & 1.785 & 2.417 & 2.332  &0.079 &0.080\\
Mg~{\it b}     & 1.577 & 2.694 & 2.651 & 3.759 & 2.927  &0.148 &0.100\\
Mg$_2$	       & 0.088 & 0.159 & 0.165 & 0.249 & 0.175  &0.015 &0.010\\
G4300	       & 2.496 & 4.656 & 4.001 & 4.819 & 4.740  &0.198 &0.100\\
C$_2$4668      & 0.322 & 1.680 & 1.906 & 4.510 & 4.382  &0.131 &0.100\\
CN$_1$	       &-0.070 & 0.025 & 0.020 & 0.063 &-0.007  &0.010 &0.005\\
CN$_2$	       &-0.047 & 0.050 & 0.047 & 0.093 & 0.018  &0.008 &0.008\\
Ca4227	       & 0.321 & 0.557 & 0.581 & 0.887 & 0.927  &0.041 &0.100\\
\enddata
\tablecomments{All indices are measured in {\AA}, except for CN$_1$ and
  CN$_2$, which are measured in magnitudes, as defined in \citet{wor94}.}
\tablenotetext{*}{Uncertainties in the Lick index measurements for
  the Milky Way globular cluster (GC) data are dominated by the
  zeropoint uncertainties in converting measured indices to the Lick
  system and are thus the same for NGC~6121, 47~Tuc, NGC~6441, and
  NGC~6528. See \citet{sch05} for details.
}
\end{deluxetable}

\begin{deluxetable}{llccccccc}
\tabletypesize{\scriptsize}
\tablecaption{EZ\_Ages Cluster Abundances Compared to Literature\label{abuntable}}
\tablewidth{0pt}
\tablehead{
\colhead{} &
\colhead{} &
\colhead{Age$^{*}$} &
\colhead{[Fe/H]} &
\colhead{[Mg/Fe]} &
\colhead{[C/Fe]} &
\colhead{[N/Fe]} &
\colhead{[Ca/Fe]} &
\colhead{Ref.}
}
\startdata
NGC 6121  &EZ\_Ages   &$7.6_{-0.7}^{+0.8}$  &$-1.30\pm0.09$ &$+0.36\pm0.11$
&$-0.12\pm0.02$  &$0.15\pm0.15$  &$-0.05\pm0.10$ &\\
          &literature &13$^{\ddag}$ &$-1.18\pm0.00$ &$+0.44\pm0.02$ &\ldots
&\ldots &$+0.26\pm0.02$ &a, b\\
\\
47 Tuc  &EZ\_Ages   &$13.9_{-3.0}^{+\mbox{max\dag}}$  &$-0.80\pm0.09$ &$+0.28\pm0.05$
&$-0.16\pm0.03$  &$+0.66\pm0.08$  &$+0.08\pm0.03$ &\\
          &literature &12 &$-0.7\pm0.05$ &$+0.4\pm0.1$ 
&$-0.2$/$0.0^{**}$ &$+1.1$/$+0.3^{**}$ &$+0.2\pm0.1$ &c, d, e\\
\\
NGC 6441  &EZ\_Ages   &$11.5_{-1.5}^{+1.9}$  &$-0.64\pm0.07$ &$+0.17\pm0.08$
&$-0.20\pm0.05$  &$+0.54\pm0.13$  &$+0.01\pm0.08$ &\\
          &literature &\ldots &$-0.43\pm0.08$ &$+0.34\pm0.09$ &\ldots &\ldots 
&$+0.03\pm0.04$ &f, g\\
\\
NGC 6528  &EZ\_Ages   &$14.6_{-2.0}^{+\mbox{max\dag}}$  &$-0.26\pm0.06$ &$+0.12\pm0.06$
&$-0.04\pm0.04$  &$+0.31\pm0.08$  &$-0.07\pm0.05$ &\\
          &literature &11 &$-0.15$/$+0.1^{\dag\dag}$
&$+0.07$/$+0.14^{\dag\dag}$ &\ldots &\ldots &$-0.40$/$+0.23^{\ddag}$ &h,
i, j, k\\
\\
M 67  &EZ\_Ages   &$4.0_{-0.58}^{+1.0}$  &$-0.09\pm0.05$ &$0.03\pm0.05$
&$-0.07\pm0.03$  &$-0.01\pm0.05$  &$-0.03\pm0.07$ &\\
          &literature &3.5 &$0.0\pm0.1$ &$0.0\pm0.1$ &$0.0\pm0.1$ 
&$0.0\pm0.1$ &$0.0\pm0.1$ &l, m, n\\
\enddata
\tablenotetext{*}{All reported ages are estimated from H$\beta$, except
  for 47 Tuc, which falls off the model grids in
  H$\beta$-$\langle$Fe$\rangle$ and is therefore calculated using
  H$\gamma_F$.}
\tablenotetext{\dag}{Positive age errors cannot be calculated by EZ\_Ages
  because they exceed the maximum 15.8 Gyr age of the models.}
\tablenotetext{\ddag}{Age determined from relative ages of NGC 6121
  and 47 Tuc in \citet{sal02}, using 47 Tuc age from \citet{sch02b}}
\tablenotetext{**}{CN-strong model / CN-weak model}
\tablenotetext{\dag\dag}{Results differ substantially between authors.
  Reported abundances are \citet{zoc04} and \citet{bar04} /
  \citealt{car01}} 
\tablerefs{(a) \citealt{sal02}, (b) \citealt{iva99}, 
  (c) \citealt{sch02a,sch02b}, (d) \citealt{bri04}, (e) \citealt{car04}, 
  (f), (g) \citet{gra06}, (h) \citealt{ort01}, 
  (i) \citealt{zoc04}, (j) \citealt{bar04}, (k) \citealt{car01}, 
  (l) \citealt{sch04a,sch04b}, (m) \citealt{tau00}, (n) \citealt{she00} }
\end{deluxetable}

\begin{deluxetable}{lcccccc}
\tabletypesize{\scriptsize}
\tablecaption{Models Using Different Indices For Fitting\label{diff_ind_tab}}
\tablewidth{0pt}
\tablehead{
\colhead{} &
\multicolumn{6}{c}{Indices used in fit} \\
\cline{2-7} \\
\colhead{Model name} &
\colhead{Balmer} &
\colhead{Fe} &
\colhead{Mg} &
\colhead{C} &
\colhead{N} &
\colhead{Ca}
}
\startdata
Standard &H$\beta$ &$\langle$Fe$\rangle$ &Mg~{\it b} &C$_2$4668 &CN$_1$ &Ca4227 \\
H$\gamma_F$ &{\bf H$\gamma_F$} &$\langle$Fe$\rangle$ &Mg~{\it b} &C$_2$4668 &CN$_1$ &Ca4227 \\
H$\delta_F$ &{\bf H$\delta_F$} &$\langle$Fe$\rangle$ &Mg~{\it b} &C$_2$4668 &CN$_1$ &Ca4227 \\
Balmer &{\bf all} &$\langle$Fe$\rangle$ &Mg~{\it b} &C$_2$4668 &CN$_1$ &Ca4227 \\
Fe4383 &H$\beta$ &{\bf Fe4383} &Mg~{\it b} &C$_2$4668 &CN$_1$ &Ca4227 \\
Fe5015 &H$\beta$ &{\bf Fe5015} &Mg~{\it b} &C$_2$4668 &CN$_1$ &Ca4227 \\
Fe5270 &H$\beta$ &{\bf Fe5270} &Mg~{\it b} &C$_2$4668 &CN$_1$ &Ca4227 \\
Fe5335 &H$\beta$ &{\bf Fe5335} &Mg~{\it b} &C$_2$4668 &CN$_1$ &Ca4227 \\
Mg$_2$ &H$\beta$ &$\langle$Fe$\rangle$ &{\bf Mg$_2$} &C$_2$4668 &CN$_1$ &Ca4227  \\
G4300  &H$\beta$ &$\langle$Fe$\rangle$ &Mg~{\it b} &{\bf G4300} &CN$_1$  &Ca4227 \\
CN$_2$ &H$\beta$ &$\langle$Fe$\rangle$ &Mg~{\it b} &C$_2$4668 &{\bf CN$_2$} &Ca4227 \\
High-z &{\bf H$\delta_F$} &{\bf Fe4383} &... &{\bf G4300} &CN$_1$ &Ca4227 \\
All    &{\bf all} &{\bf all} &{\bf all} &{\bf all} &{\bf all}     &Ca4227 \\
\enddata
\tablecomments{For indices used in the fitting process, bold face
  indicates the index that differs from the standard set.  ``All''
  indicates that all available indices for that element were averaged
  together (i.e., fitting for [C/Fe] was done using an average of
  C$_2$4668 and G4300).  
}
\end{deluxetable}

\begin{deluxetable}{lccccccccccccc}
\tabletypesize{\scriptsize}
\tablecaption{EZ\_Ages Results Using Different Indices For Fitting\label{diff_fit_tab}}
\tablewidth{0pt}
\tablehead{
\colhead{} &
\multicolumn{6}{c}{EZ\_Ages results for NGC~6441} &
\colhead{} &
\multicolumn{6}{c}{EZ\_Ages results for M~67} \\
\cline{2-7} \cline{9-14} \\
\colhead{Model} &
\colhead{Age} &
\colhead{[Fe/H]} &
\colhead{[Mg/Fe]} &
\colhead{[C/Fe]} &
\colhead{[N/Fe]} &
\colhead{[Ca/Fe]} &
\colhead{} &
\colhead{Age} & 
\colhead{[Fe/H]} &
\colhead{[Mg/Fe]} &
\colhead{[C/Fe]} &
\colhead{[N/Fe]} &
\colhead{[Ca/Fe]}
}
\startdata
Standard    &11.4 &$-$0.64 &$+$0.17 &$-$0.20 &$+$0.62 &$+$0.04 & &4.0 &$-0.09$ &$+0.03$ &$-0.07$ &$-0.01$ &$-0.03$\\
H$\gamma_F$  &6.9 &$-$0.54 &$+$0.17 &$-$0.22 &$+$0.60 &$+$0.05 & &3.5 &$-0.07$ &$+0.03$ &$-0.06$ &$-0.01$ &$+0.01$\\
H$\delta_F$  &8.2 &$-$0.57 &$+$0.17 &$-$0.22 &$+$0.60 &$+$0.05 & &3.7 &$-0.08$ &$+0.05$ &$-0.06$ &$-0.01$ &$-0.04$\\
Balmer       &8.3 &$-$0.57 &$+$0.17 &$-$0.21 &$+$0.60 &$+$0.05 & &3.7 &$-0.08$ &$+0.03$ &$-0.06$ &$-0.01$ &$-0.01$\\
Fe4383      &11.1 &$-$0.55 &$+$0.08 &$-$0.25 &$+$0.58 &$-$0.04 & &3.8 &$-0.02$ &$-0.03$ &$-0.10$ &$-0.03$ &$-0.09$\\
Fe5015$^{\dag}$&...&...      &...      &...     &...      &... & &4.3 &$-0.17$ &$+0.08$ &$-0.02$ &$-0.04$ &$+0.01$\\
Fe5270      &11.3 &$-$0.62 &$+$0.15 &$-$0.21 &$+$0.60 &$+$0.01 & &3.9 &$-0.10$ &$+0.03$ &$-0.07$ &$-0.01$ &$-0.03$\\
Fe5335      &11.5 &$-$0.66 &$+$0.19 &$-$0.19 &$+$0.64 &$+$0.06 & &4.0 &$-0.09$ &$+0.03$ &$-0.07$ &$-0.01$ &$-0.03$\\
Mg$_2$      &11.4 &$-$0.63 &$+$0.22 &$-$0.20 &$+$0.60 &$+$0.02 & &4.0 &$-0.09$ &$+0.04$ &$-0.07$ &$-0.01$ &$-0.03$\\
G4300$^{*}$ &11.6 &$-$0.65 &$+$0.15 &$-$0.33 &$+$0.79 &$+$0.02 & &...  &...      &...      &...  &...     &...    \\
CN$_2$      &11.4 &$-$0.64 &$+$0.16 &$-$0.20 &$+$0.56 &$+$0.02 & &4.0 &$-0.09$ &$+0.03$ &$-0.07$ &$-0.03$ &$-0.04$\\
High-z$^{*}$&5.7  &$-$0.34 &$+$0.03 &$-$0.33 &$+$0.50 &$-$0.10 & &2.7 &$+0.15$ &...     &$-0.15$ &$-0.04$ &$-0.13$\\
All$^{*\dag}$&7.5 &$-$0.47 &$+$0.12 &$-$0.26 &$+$0.53 &$-$0.02 & &3.7 &$-0.07$ &$+0.03$ &$-0.06$ &$-0.02$ &$-0.01$\\
\enddata
\tablenotetext{*}{There is a problem with G4300 in the M~67 spectrum
  \citep{sch04a}, thus for this cluster no separate model for G4300 is
  computed and the ``high-z'' and ``all'' models are computed with
  C$_2$4668.}
\tablenotetext{\dag}{Fe5015 falls on a bad CCD column in the spectrum
  of NGC~6441, thus for this cluster no separate model for Fe5015 is
  computed and the ``all'' model does not include Fe5015.  }
\end{deluxetable}

\clearpage

\clearpage

\end{document}